\documentclass[20pt.,preprint]{emulateapj}
\usepackage{graphics}

\shorttitle{Non-linear CRCD Instability}
\shortauthors{Riquelme \& Spitkovsky}

\begin{document}
\title{Non-linear Study of Bell's Cosmic Ray Current-driven Instability}

\author{Mario A. Riquelme and Anatoly Spitkovsky}

\affil{Department of Astrophysical Sciences, Princeton University, Princeton, NJ 08544}

\email{marh@astro.princeton.edu, anatoly@astro.princeton.edu}

%\altaffiltext{1}{Departamento de Astronomia y Astrofisica, P. Universidad Catolica de Chile, where part of this work was realized.}

%\email{marh@astro.princeton.edu, dns@astro.princeton.edu}

%\altaffiltext{1}{Departamento de Astronomia y Astrofisica, P. Universidad Catolica de Chile, where part of this work was realized.}

\begin{abstract}
The cosmic ray current-driven (CRCD) instability, predicted by \cite{Bell04}, consists of non-resonant, growing plasma waves driven by the electric current of cosmic rays (CRs) that stream along the magnetic field ahead of both relativistic and non-relativistic shocks. Combining an analytic, kinetic model with one-, two-, and three-dimensional particle-in-cell simulations, we confirm the existence of this instability in the kinetic regime and determine its saturation mechanisms. In the linear regime, we show that, if the background plasma is well magnetized, the CRCD waves grow exponentially at the rates and wavelengths predicted by the analytic dispersion relation. The magnetization condition implies that the growth rate of the instability is much smaller than the ion cyclotron frequency. 
As the instability becomes non-linear, significant turbulence forms in the plasma. This turbulence reduces the growth rate of the field and damps the shortest wavelength modes, making the dominant wavelength, $\lambda_d$, grow proportional to the square of the field. At constant CR current, we find that plasma acceleration along the motion of CRs saturates the instability at the magnetic field level such that $v_{A} \sim v_{d,cr}$, where $v_{A}$ is the Alfv\'en velocity in the amplified field, and $v_{d,cr}$ is the drift velocity of CRs. The instability can also saturate earlier if CRs get strongly deflected by the amplified field, which happens when their Larmor radii get close to $\lambda_d$. We apply these results to the case of CRs propagating in the upstream medium of the forward shock in supernova remnants. If we consider only the most energetic CRs that escape from the shock, we obtain that the field amplification factor of $\sim 10$ can be reached. This confirms the CRCD instability as a potentially important component of magnetic amplification process in astrophysical shock environments.
\end{abstract}


\section{Introduction}
High energy particle acceleration and magnetic field amplification appear to be tightly related phenomena in many astrophysical environments. For instance, X-ray observations of synchrotron emission from ultrarelativistic electrons in young supernova remnants (SNRs) \citep{VolkEtAl05, Ballet06, UchiyamaEtAl07} suggest that electrons are efficiently accelerated in these environments, and that the ambient magnetic field in the dowstream medium of SNR forward shocks is amplified by a factor of $\sim 100$ compared to its typical value in the interstellar medium. Also, cosmic rays (CRs) with energies up to $\sim 10^{15}$eV are believed to originate in SNRs. Calculations based on the diffusive shock acceleration mechanism \citep{Krymsky77, AxfordEtAl77, Bell78, BlandfordEtAl78} show that such high energies can only be reached if CRs are efficiently confined to the remnant \citep{LagageEtAl83}, a condition that would be eased by a substantial magnetic field amplification.%\newline

While it is supported by direct observations of SNRs, the process by which the field amplification takes place is still a mystery. \cite{Bell04} suggested the possibility of magnetic amplification driven by non-resonant CRs propagating in the upstream region of shocks. This instability is different from the Alfv\'en wave amplification due to cyclotron resonance with CRs \citep{KulsrudEtAl69, McKenzieEtAl82}. Instead, it only requires positively charged CRs propagating along a background magnetic field, $\vec{B}_0$, with Larmor radii much larger than the wavelength of the wave. This condition is expected to be satisfied by diffusively shock-accelerated CRs in the upstream region of both relativistic and non-relativistic shocks. While individual CRs are relativistic, on average they stream with respect to the upstream plasma with a drift velocity, $\vec{v}_{d,cr}$, that depends on the distance from the shock. The lowest energy CRs are efficiently confined to the shock vicinity by the upstream turbulence, and drift with the shock with $\vec{v}_{d,cr} \approx \vec{v}_{sh}$, where $\vec{v}_{sh}$ is the shock speed. The higher energy CRs are less affected by the upstream scattering, so they tend to escape more easily from the shock. Thus, as the distance from the shock increases, $\vec{v}_{d,cr}$ will increase, asymptotically approaching $c/2$ for non-relativistic shocks (for relativistic shocks, $\vec{v}_{d,cr} \approx c$ everywhere). Also, considering that electrons are more affected by radiative losses than ions, it is reasonable to think that at some distance from the shock CRs will be mainly positively charged particles. Their drift will drive a constant current, $\vec{J}_{cr}$, through the upstream plasma. This current will be compensated by the opposite return current, $\vec{J}_{ret} \approx -\vec{J}_{cr}$, provided by the background plasma. If there is a small magnetic field perturbation $\vec{B}_{tr}$, perpendicular to $\vec{B}_0$, a $\vec{J}_{ret} \times \vec{B}_{tr}$ force will push the background plasma transversely. A force of the same magnitude will push the CRs in the opposite direction, so that the net force acting on the background plasma-CRs system vanishes. However, given the high rigidity of the CRs, only the background plasma will experience a significant transverse motion. For a helical magnetic perturbation $\vec{B}_{tr}$, this transverse motion will stretch the magnetic field lines, producing an amplification of $\vec{B}_{tr}$. Using analytic MHD analysis, \cite{Bell04} showed that for a right-handed circularly polarized electromagnetic wave, this amplification would be exponential, and faster than the resonant instability\footnote{As mentioned by \cite{Bell04}, the polarization of the waves will be right-handed when $\vec{B}_0$ and $\vec{J}_{cr}$ are parallel. In the antiparallel case, the polarization is left-handed}.

The linear dispersion relation of this cosmic ray current-driven (CRCD) instability has been calculated using both MHD \citep{Bell04} and kinetic treatments \citep{RevilleEtAl06, BlasiEtAl07}. These works found that, if $\vec{J}_{cr}$ is kept constant, the instability will grow at a prefered wavelength $\lambda_{max} = B_0c/J_{cr}$, with a growth rate $\gamma_{max} = J_{cr}(\pi/\rho c^2)^{1/2}$, where $\rho$ is the mass density of the background plasma.

The non-linear evolution of the CRCD instability has been studied making use of both MHD and particle-in-cell (PIC) simulations. The MHD studies \citep{Bell04, Bell05, ZirakashviliEtAl08} have shown a substantial amplification of the ambient magnetic field and the formation of turbulence, which is characterized by prominent density fluctuations in the plasma. They have established a saturation criterion that depends on the wavelength, $\lambda$, of each mode and that is given by $B_{sat}(\lambda) \sim  B_0 \lambda/\lambda_{max}$. This saturation criterion would imply that, if $J_{cr}$ is constant, the field never stops growing but only migrates into longer wavelengths. This migration would be such that the dominant wavelength, $\lambda_d$, goes roughly as $\lambda_d \sim \lambda_{max}(B/B_0)$, where B is the mean value of the magnetic field \citep{Bell04}. The saturation in MHD simulations with constant CR current is either not observed \citep{ZirakashviliEtAl08} or is due to the size of magnetic fluctuations reaching the size of the box \citep{Bell04}. 

One important limitation of the MHD simulations is that they cannot follow the instability at arbitrarily low densities, which makes it difficult to model large plasma density fluctuations properly. Also, the MHD simulations do not include the back-reaction on the CRs, which must play a fundamental role in the saturation of the instability. Both difficulties can be potentially resolved by fully kinetic PIC simulations.

A first attempt to use PIC simulations for this problem was made by \cite{NiemiecEtAl08}. Even though their results show magnetic amplification, it accurs at a significantly lower rate and through a kind of turbulence that essentially differs from the circularly polarized, growing waves predicted by \cite{Bell04}. This raised a question about the existence of the CRCD instability beyond the MHD approximation.

In this work we confirm the existence of the CRCD instability using PIC simulations and establish the conditions under which it is present. In order to understand the saturation mechanisms of the instability, we separate our study in three parts, presented in \S\ref{sec:thewaves}, \S\ref{sec:multidimensional}, and \S\ref{freecrs}. In \S\ref{sec:thewaves}, we show the main non-linear properties of the CRCD waves, focusing on their intrinsic saturation mechanism in the presence of constant CR current, $\vec{J}_{cr}$. We do this using an one-dimensional, analytic model, and check our results with one-dimensional PIC simulations. We calculate a non-linear dispersion relation that includes the time evolution of the phase velocity of the waves. Also, our model quantifies all the plasma motions induced by the waves, which is needed to understand the wave behavior in the multidimensional context. In \S\ref{sec:multidimensional}, we present the multidimensional properties of the instability using two- and three-dimensional simulations, also with constant $\vec{J}_{cr}$. First, we determine the conditions under which the CRCD instability can grow without being affected by plasma filamentation as in the case of \cite{NiemiecEtAl08}. Then, combining the multidimensional simulations with our results from \S \ref{sec:thewaves}, we determine the main properties of the instability in its non-linear stage. We confirm the generation of turbulence as suggested by previous MHD studies, and reexamine its main properties. We estimate the typical turbulence velocity and length scale as a function of the magnetic amplification, finding a faster migration to longer wavelengths than predicted by MHD simulations. We find that the acceleration of background plasma along the direction of motion of the CRs causes the intrinsic saturation of the CRCD instability at constant $\vec{J}_{cr}$. 
In \S \ref{freecrs}, we study the effect of the back-reaction on the CRs as a second saturation mechanism for the instability. \S \ref{conclusions} presents our conclusions and an application to the case of SNR environments.

\section{CRCD waves}\label{sec:thewaves}
In this section we present the one-dimensional analysis of the CRCD waves at constant $\vec{J}_{cr}$, i.e., without considering the back-reaction on the CRs. In \S2.1 we show an analytic, kinetic model for the CRCD waves, valid in the non-linear regime. After that, in \S2.2, we check our model making use of
one-dimensional PIC simulations. We consider a piece of upstream plasma through which positively charged CRs flow, providing the current $\vec{J}_{cr}$. We focus on the situation where the initial magnetic field $\vec{B}_0$, $\vec{J}_{cr}$, and the wave vector of the CRCD mode $\vec{k}$, are parallel. Local charge neutrality is assumed as initial condition, so $n_i + n_{cr} = n_e$, where $n_i$, $n_e$, and $n_{cr}$ correspond to the density of ions, electrons, and CRs, respectively\footnote{This charge neutrality implies that the background plasma must extract a small amount of additional electrons from its surroundings to compensate the CR charge.}.  %parallel to the initial magnetic field, 

\subsection{Analytic, kinetic model}\label{sec:analytical}
 
In this section we study the time evolution of CRCD waves of different wavevectors $\vec{k}$, which are characterized by their growth rate $\gamma$ and phase velocity $\omega/k$, where $k = |\vec{k}|$. The derivation is made for a right-handed polarized wave, and is based on the calculation of the drift velocities of plasma components in the presence of a wave of arbitrary amplitude. If we know these drift velocities and the number densities of the different species, we can calculate the total current provided by the background plasma as a function of space and time. Adding this total plasma current to the constant $\vec{J}_{cr}$ contributed by the CRs, the time evolution of the wave can be directly obtained from the Ampere's and Faraday's laws. The details of the calculation are presented in Appendix \ref{app:analytic}. Here we describe its main results, which are summarized in the dispersion relation given by Equation (\ref{eq:dispersion}). This dispersion relation assumes a constant $\gamma$ and allows $\omega/k$ to evolve in time. We will see below that $\gamma$ is indeed constant as long as $v_{d,cr} \gg v_{A}$, where $v_{A}$ is the Alfv\'en velocity of the backgroud plasma. This condition not only puts a limit to the validity of the derivation, but also sets a saturation criterion for the CRCD waves. Also, our derivation is in the low plasma temperature limit and assumes that $v_{A,0} \gg (n_{cr}/n_{i})v_{d,cr}$, where $v_{A,0}$ is the initial Alfv\'en velocity of the plasma. Using two-dimensional PIC simulations, we show below that this second condition is actually a requirement for the CRCD waves not to be quenched by Weibel-like plasma filamentation.
From the real part of Equation (\ref{eq:dispersion}), we see that the growth rate, $\gamma$, is
maximized when the wavenumber $k=k_{max}=2\pi J_{cr}/B_0 c$, which corresponds to the same wavenumber of maximum growth found in the linear regime \citep{Bell04, RevilleEtAl06, BlasiEtAl07}. From the imaginary part, we obtain the following differential equation for $\omega$ as
a function of the amplification factor of the waves, $f$ (defined as the ratio between the magnitude of the transverse magnetic field, $B_{tr}$, and $B_0$), 
\begin{equation}
%\begin{array}{rcl}
\frac{f}{2}\bigg(1+\frac{c^2}{v_A^2}\bigg)\frac{d\omega}{df} + \omega
\bigg(1+\frac{c^2}{v_A^2}\frac{1}{1+f^2}\bigg) - k_{max}\frac{c^2}{v_{d,cr}} = 0.
%\end{array}
\label{eq:diferencial}
\end{equation}
The solution for Equation (\ref{eq:diferencial}) is
\begin{equation}
\omega = \frac{k_{max}c^2}{\Big(1+\frac{c^2}{v_A^2}\Big)v_{d,cr}}, 
\label{eq:phasevelocity}
\end{equation}
which, when $v_A \ll c$, can be approximated as $\omega \approx k_{max}v_A^2/v_{d,cr}$. This implies that, although in the linear regime the CRCD waves are almost purely growing ($\omega/k_{max} \ll v_{A,0}$), if $v_A$ gets close to $v_{d,cr}$, their phase velocity can also become comparable to $v_{d,cr}$.

Taking the real part of Equation (\ref{eq:dispersion}) and evaluating at $k =
k_{max}$, we obtain that
\begin{equation}
\gamma(k_{max})^2 \equiv \gamma_{max}^2 \approx
k_{max}^2v_{A,0}^2\bigg(1-2\frac{v_A^2}{c^2}\bigg)\bigg(1-2\frac{v_A^2}{v_{d,cr}^2}\bigg),
\label{eq:growthrate}
\end{equation}
where we have kept terms only to first order in $v_A^2/v_{d,cr}^2$ and $v_A^2/c^2$.
We see that in the regime $v_A \ll v_{d,cr}$, the instability grows exponentially
with a maximum growth rate, $\gamma_{max} \approx k_{max}v_{A,0}$ that is constant and has the same value as obtained in the previous linear studies \citep{Bell04,
RevilleEtAl06, BlasiEtAl07}. Even though Equation (\ref{eq:growthrate}) shows that our assumption of constant $\gamma_{max}$ is only valid when $v_A \ll v_{d,cr}$, it also indicates that, as $v_A$ approaches $v_{d,cr}$, the growth rate will be substantially reduced, suggesting an intrinsic saturation limit for the CRCD waves at $v_A \sim v_{d,cr}$.

In Appendix \ref{app:analytic} we also show that the presence of the CRCD waves induces bulk motions of the plasma particles  both parallel and transverse to $\vec{J}_{cr}$. The parallel motion has a velocity $v_{x,an} \approx f^2v_{A,0}^2/v_{d,cr}$, while the transverse motion has a velocity $v_{tr,an} \approx f v_{A,0}$, which always points perpendicular to $\vec{B}_{tr}$ (and to $\vec{J}_{cr}$). The parallel plasma motion implies that, when $v_A \sim v_{d,cr}$, the entire plasma will move at a speed close to $v_{d,cr}$, which, from the point of view of the plasma, substantially reduces $\vec{J}_{cr}$. This reduction in $\vec{J}_{cr}$ explains the intrinsic saturation of the waves at $v_A \sim v_{d,cr}$. The tranverse motion, on the other hand, becomes of the order of the Alfv\'en velocity of the plasma when the CRCD waves become non-linear ($f \gtrsim 1$). We will see below that this increasing transverse velocity is related to turbulence formation in the non-linear regime.

\subsection{One-dimensional Simulations}\label{1Dsimu}
We checked our analytical results with one-dimensional PIC simulations.  We use the PIC code TRISTAN-MP \citep{Buneman93,Spitkovsky05}, which can run in one, two, and three dimensions. In these simulations, like in our analytic model, all plasma properties depend only on one spatial direction ($x$), but both the velocities of the particles and the electromagnetic fields keep their three-dimensional components. We set up a periodic box that contains an initially cold background plasma (with typical particle thermal velocity of $10^{-4}c$) composed of ions and electrons, and a small population of relativistic ions (CRs). The driving current is given by the CRs that move along $\hat{x}$ with a mean velocity $v_{d,cr}$ that we vary between runs. These CRs are not allowed to change their velocities, as if they had an infinite Lorentz factor, $\Gamma$. Such ``locked" CRs allow us to study the non-linear evolution of the instability considering a constant $\vec{J}_{cr}$, i.e., eliminating the back-reaction on the CRs. We give electrons a small velocity along $\hat{x}$ such that the background plasma carries a current $-J_{cr}\hat{x}$. This way the net current is zero. The initial magnetic field, $\vec{B}_0$ also points along $\hat{x}$. Since we want to simulate a situation where $n_{cr} \ll n_i$, having good CR statistics would imply a large number of particles per cell. In order to overcome this difficulty, we have initialized the same number of macroparticles for CRs, ions, and electrons, but modified their charges so that $q_i = -q_e(1-\alpha)$ and $q_{cr} = -q_e\alpha$, where $\alpha \equiv n_{cr}/n_i$. We change the mass of the particles accordingly in order to keep the right charge to mass ratios. Particles are initially located randomly in the box such that at the position of each ion we also have an electron and a CR, so the initial charge density is zero. This initialization is also used in our two- and three-dimensional runs. The common numerical parameters for the simulations are $c/\omega_{p,e} \approx 3.3$ $\Delta$ (where $\omega_{p,e}$ is the electron plasma frequency and $\Delta$ is the grid cell size), ion-electron mass ratio $m_i/m_e = 100$, speed of light $c=0.1125$ $\Delta$/$\Delta_t$ (where $\Delta_t$ is the time step), and 12.5 particles per species per cell.

\subsubsection{Relativistic Regime}\label{sec:rel}
\begin{figure*}
\begin{center}
\centering\includegraphics[width=13cm]{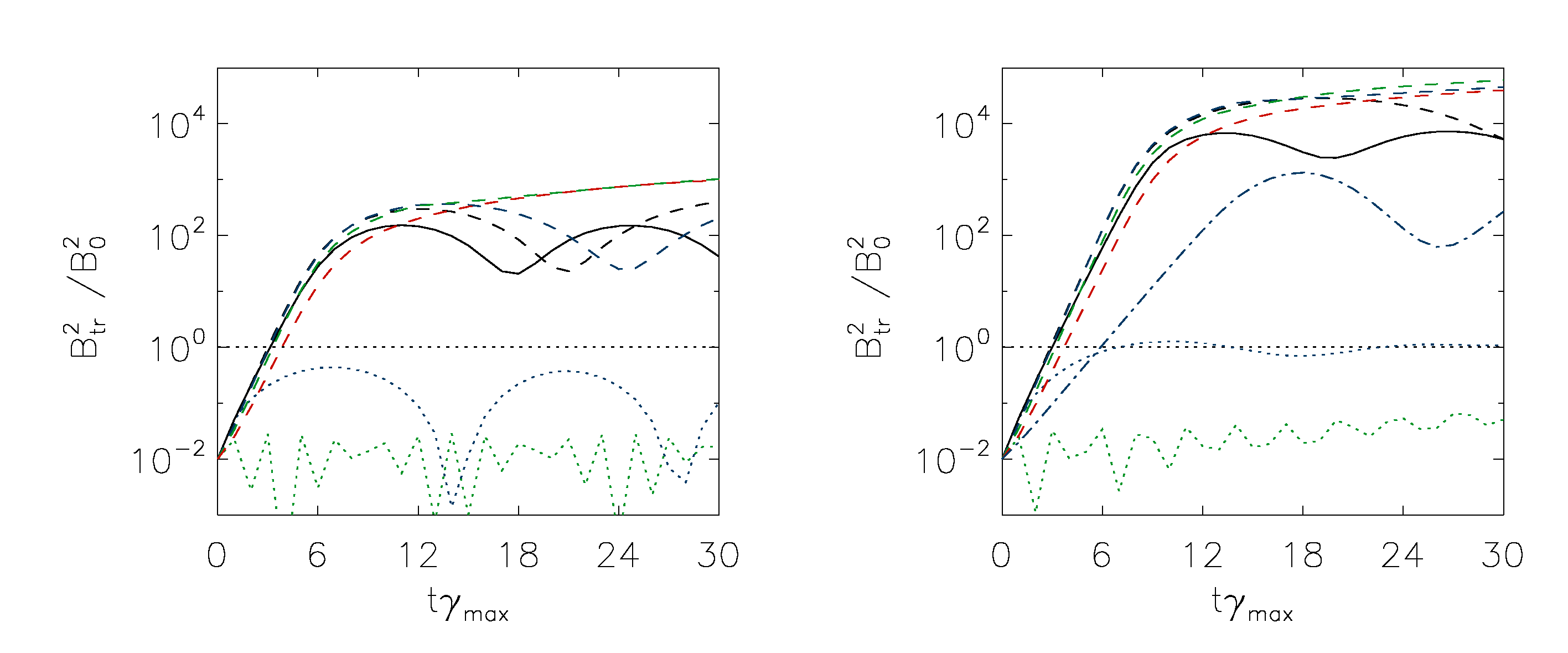}
\caption{Magnetic energy growth, $B_{tr}^2/B_{0}^2$, is plotted as a function of time, $t$, for one-dimensional runs with constant $\vec{J}_{cr}$. All the runs (except for the one represented by the blue, dot-dashed line in the right panel) are relativistic ($v_{d,cr} \cong c$). In each
case we use  as a seed a single CRCD wave whose wavelength, $\lambda$, is the size of the simulation box, and the initial amplitude is $0.1B_{0}$. In all the relativistic cases the density of CRs is
such that the theoretical maximum growth rate, $\gamma_{max} = 0.2 \omega_{c,i}$, so $\gamma_{max} \ll \omega_{c,i}$. $v_{A,0}/c = 1/10$ for the runs in the left panel and 1/80 for the ones in the right panel. The green and blue dotted lines correspond to modes with wavelength $\lambda = $0.5 and $0.75\lambda_{max}$, the black solid line to $\lambda=\lambda_{max}$, and the black, blue, green, and red dashed lines correspond to $\lambda = 1.25, 1.5, 2$, and 3$\lambda_{max}$, where $\lambda_{max}$ is the wavelength of the theoretically determined fastest growing mode. The black dotted line represents $B_x^2$ in all cases. The blue dot-dashed line in the right panel corresponds to a non-relativistic counterpart of the $v_{A,0}/c = 1/80$, $\lambda = 1.5\lambda_{max}$ simulation (which is the one that grows the fastest), i.e., they have the same parameters except $v_{d,cr}/c = 0.5$ and $\lambda = 3\lambda_{max}$. The results are consistent with the theoretical $\gamma_{max}$ and $\lambda_{max}$, and with the intrinsic saturation criterion, $v_A \sim v_{d,cr}$.} \label{fg:figura1}
\end{center}
\end{figure*}

Here we test the non-linear dispersion relation found in \S \ref{sec:analytical} in the relativistic regime ($v_{d,cr} \cong c$), which would be appropriate for the upstream medium of a relativistic shock front. We ran one-dimensional simulations in boxes of different sizes $L$,
set up to probe the growth rate of different wavelengths, $\lambda$. As an
initial condition we used a right-handed, circularly polarized, growing wave of amplitude 0.1$B_{0}$, whose fields and particle velocities were determined from our analytic model (Appendix \ref{app:analytic}). We put only one period of the wave in a box, so that the only other modes that could be excited are shorter or equal to $L/2$. We choose the density of CRs such that the corresponding maximum growth rate of the instability, $\gamma_{max}$, is 0.2 $\omega_{c,i}$, so we are in the regime where the background plasma is well magnetized. Simulations were run for $\lambda$ equal to 0.5, 0.75, 1, 1.25, 1.5, 2, and 3$\lambda_{max}$. We used several values for the initial Alfv\'en velocities, $v_{A,0}/v_{d,cr}$, in the range $1/10$ to $1/80$. This implies that the initial gyrotime $2\pi \omega_{c,e}^{-1}$ ranges from 189 to 1456 $\Delta_t$, so it is resolved with about 20 $\Delta_t$ even when $v_{A} \sim v_{d,cr}$, and $\omega_{p,e}/\omega_{c,e}$ ranges from 1 to 8. The results for the cases $v_{A,0}/v_{d,cr} = 1/10$ and $1/80$ are presented in Fig. \ref{fg:figura1}.
We observe that for $L = 0.5\lambda_{max}$ there is practically no growth.
For $L=$ 1, 1.25, and 1.5$\lambda_{max}$ we obtain nearly the
same growth rate, which is very close to the analytic $\gamma_{max}$. For longer wavelengths, the growth rate
gradually decreases. In all our experiments, for $L \ge \lambda_{max}$, the
exponential growth continues until $v_A$ becomes close to $c$, which confirms our analytical saturation criterion for fixed CRs. At later times we see that, depending on $L$, the amplitude of the wave either oscillates or keeps growing but at a much lower rate. 

\subsubsection{Non-relativistic Regime}\label{sec:nonrel}
While individual CRs near a non-relativistic shock move at almost the speed of light, on average they move with respect to the upstream at a drift velocity, $v_{d,cr}$, that is less than c. In order to study this case, we ran a series of simulations where, besides not allowing CRs to alter their trajectories, we make them drift along $\hat{x}$ at a velocity $v_{d,cr} = 1, 0.9, 0.8, 0.6, 0.4,$ and $0.2 c$. We do this using a box size $L=20\lambda_{max}$. In this case we do not seed the instability with a small amplitude, growing wave, as done in \S \ref{sec:rel}. Instead, we only put the initial magnetic field, $\vec{B}_0$, such that $v_{A,0}/c = 1/10$, forming an angle $\theta = 5^o$ with $\vec{J}_{cr} = J_{cr}\hat{x}$. We use this set-up to show that the instability can develop from any kind of noise. We tilt $\vec{B}_0$ by a small angle to inject a small amount of magnetic and kinetic energy perpendicular to $\vec{J}_{cr}$, so it acts as an initial seed that does not favor any particular $\lambda$ (experiments with $\theta = 0$ were also run, showing no difference besides requiring a longer initial time for the wave to appear). Other numerical parameters are the same as in the relativistic experiments.
\begin{figure}
\centering\includegraphics[width=7.75cm]{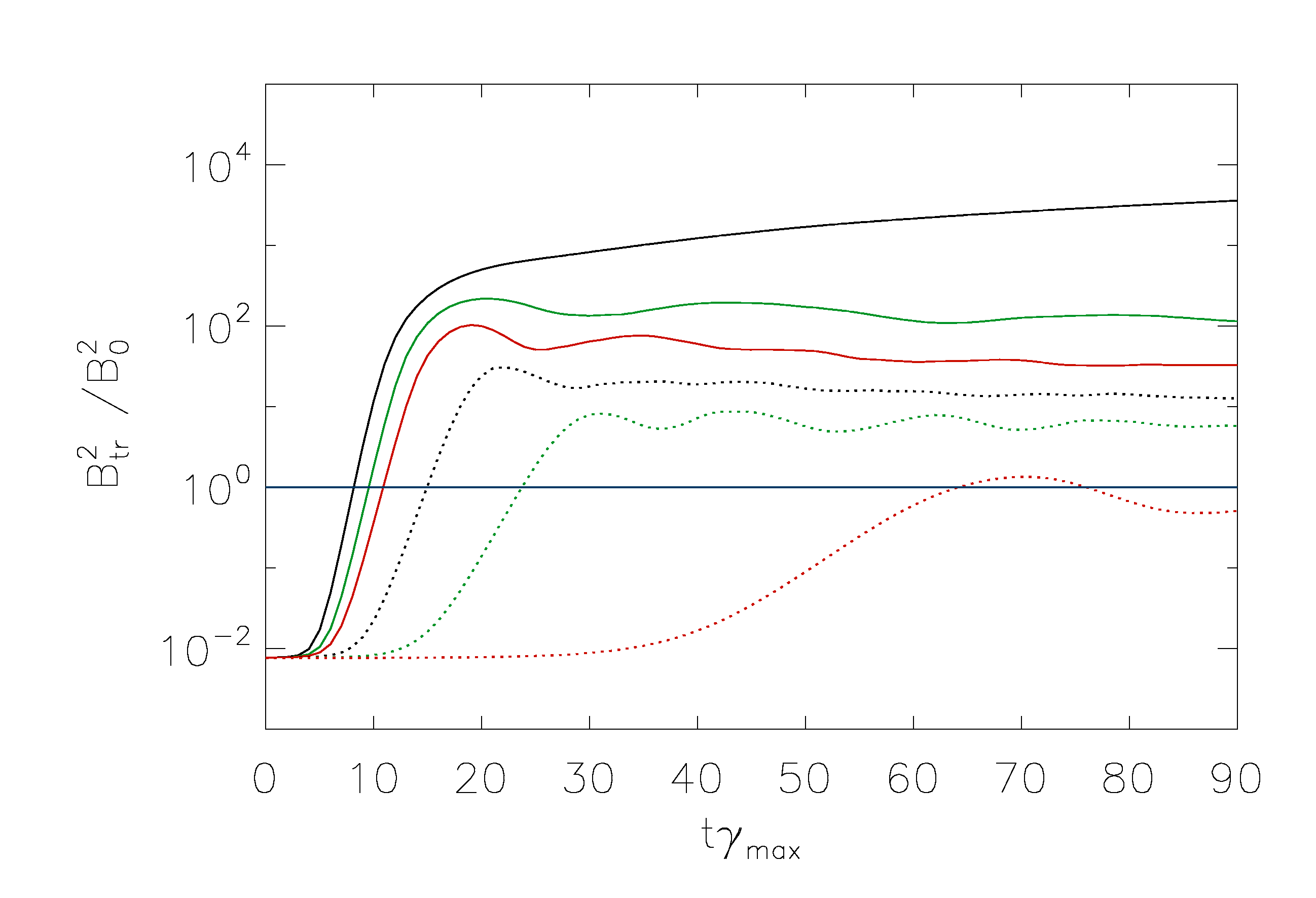}
\caption{$B_{tr}^2/B_0^2$ is plotted as a function of time, $t$, for experiments similar to the ones depicted in Fig. \ref{fg:figura1}, but for $v_{A,0}/c = $ 1/10 and a box of $20\lambda_{max}$, where $\lambda_{max}$ is the wavelength of the theoretically determined fastest growing mode. The instability is seeded by tilting $\vec{B}_0$ by $\sim 5^o$ with respect to $\vec{J}_{cr}$. A series of $v_{d,cr}$ is tested: $v_{d,cr}/c = 1$ (solid black), 0.9 (solid green), 0.8 (solid red), 0.6 (dotted black), 0.4 (dotted green), and 0.2 (dotted red). $\gamma_{max}$ is the maximum theoretical growth rate for the case $v_{d,cr}/c = 1$. The results are consistent with the theoretical $\gamma_{max}$ and $\lambda_{max}$, and with the intrinsic saturation criterion, $v_A \sim v_{d,cr}$.} \label{fg:lesscurr}
\end{figure}
Fig. \ref{fg:lesscurr} shows the magnetic energy evolution for the six $v_{d,cr}$ tested. We observe that the growth rate is $\gamma \approx \gamma_{max}v_{d,cr}/c$, where $\gamma_{max}$ is the maximum growth rate for $v_{d,cr} = c$. The amplitude at which the exponential growth stops is such that $v_A \sim v_{d,cr}$, confirming our results from \S \ref{sec:analytical}. 
\begin{figure}
\centering\includegraphics[width=8.5cm]{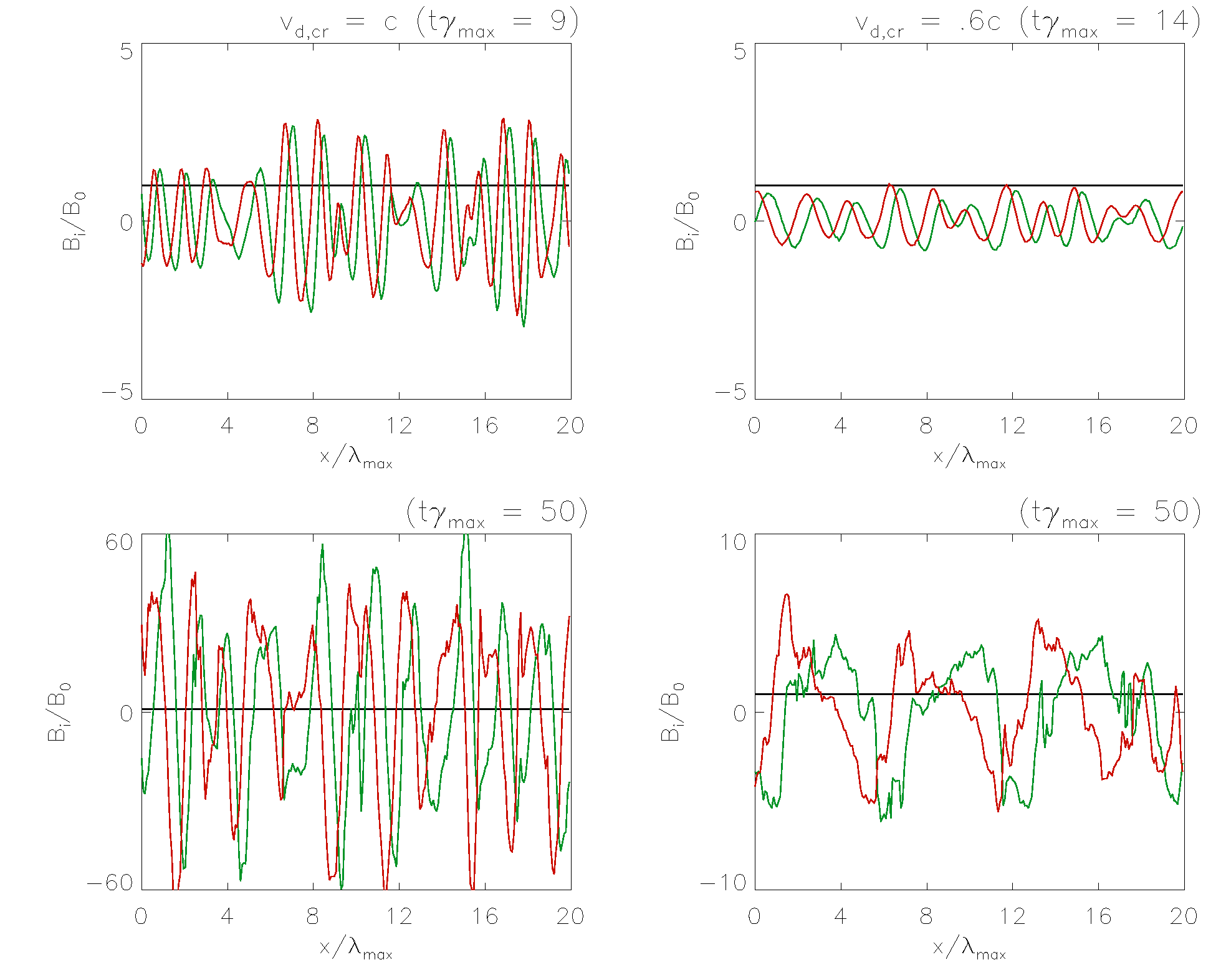}
\caption{Transverse components of the magnetic field, $B_y$(red line) and $B_z$(green line), are plotted as a function of distance, $x$, at two different times for two of the runs described in Fig. \ref{fg:lesscurr} ($v_{d,cr}/c = 1$ and 0.6). The black line represents $B_x$. 
The two left plots show the case $v_{d,cr}/c=1$ at $t\gamma_{max} = 9$ and 50. The two right plots show the case $v_{d,cr}/c = 0.6$ at $t\gamma_{max} = 14$ and 50. The results are consistent with the instability appearing initially as right-handed circularly polarized wave with a preferred wavelength $\sim \lambda_{max}(v_{d,cr}/c)$, and with a migration into longer wavelengths as the instability grows.}\label{fg:sequenceapj}
\end{figure}
Fig. \ref{fg:sequenceapj} shows the different components of the magnetic field as a function of position, $x$, at different times for $v_{d,cr} = c$, and $v_{d,cr} = 0.6 c$. We can see that in both cases the instability appears as a right-handed polarized wave and at an initial wavelength $\lambda \approx \lambda_{max}c/v_{d,cr}$, where $\lambda_{max}$ is the wavelength of maximum growth for $v_{d,cr} = c$. After the wave reaches saturation, there is a migration into longer wavelengths. This migration appears because the modes with wavelengths greater than $\lambda_{max}$ grow more slowly, but still grow and saturate at $v_A \sim v_{d,cr}$, as can be seen in Fig. \ref{fg:figura1} for the relativistic regime ($v_{d,cr} = c$). It means that, as the instability reaches $v_A \sim v_{d,cr}$, the spectrum of the waves gradually receives more contribution from wavelengths longer than $\lambda_{max}$.

\subsubsection{Background plasma motion}\label{sec:motion}
\begin{figure*}
\centering\includegraphics[width=16cm]{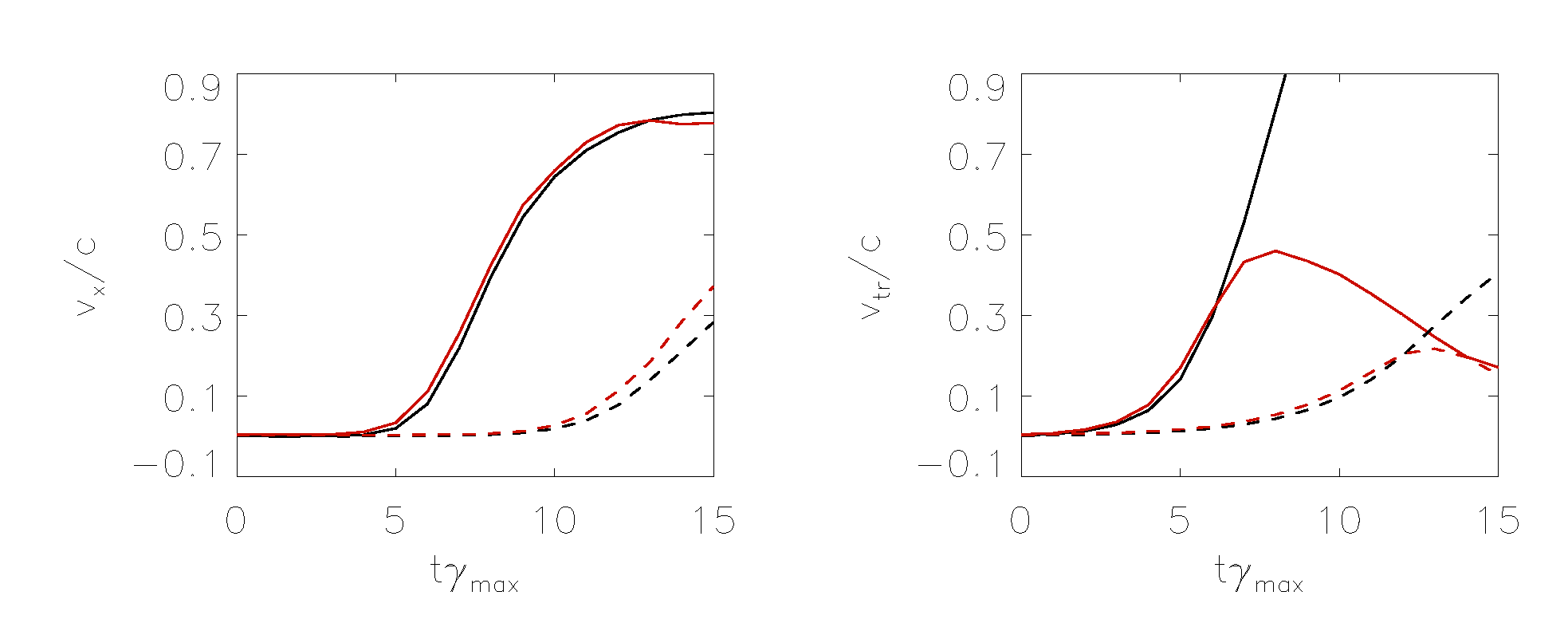}
\caption{The mean velocity along $\hat{x}$, $v_{x}$ (left panel), and the mean magnitude of the transverse velocity, $v_{tr}$ (right panel), as a function of time.  Two simulations described in the right panel of Fig. \ref{fg:figura1} ($v_{A,0}/c = 1/80$) are displayed, namely, a simulation with $\lambda=1.5 \lambda_{max}$ for $v_{d,cr}/c = 1$ (solid lines) and a simulation with $\lambda=3 \lambda_{max}$ for $v_{d,cr}/c = 0.5$ (dashed lines). The velocities are represented using red lines. For comparison, our analytic estimates for the plasma velocities, $v_{x,an} = f^2v_{A,0}^2/v_{d,cr}$ and $v_{tr,an} = fv_{A,0}$, are shown in black lines. As seen in the rigth panel of Fig. \ref{fg:figura1}, saturation for the $v_{d,cr}/c = 1$ and 0.5 runs happens at $t\gamma_{max} = 7$ and 12, respectively. Our analytic estimates for the plasma velocities are, therefore, consistent with our numerical results.}\label{parperfig}
\end{figure*}

As we saw in \S \ref{sec:analytical}, CRCD waves induce plasma motions both parallel and perpendicular to $\hat{x}$. The left panel in Fig. \ref{parperfig} shows the mean velocity of plasma particles along $x$ (i.e., parallel to $\vec{J}_{cr}$), and the analytic estimate for this velocity, $\vec{v}_{x,an} \approx f^2v_{A,0}^2/v_{d,cr}$. This velocity comes from the $\vec{E} \times \vec{B}$ drift of background particles, which in a well magnetized plasma ($\gamma_{max} \ll \omega_{c,i}$) is much larger than other plasma drifts (see Appendix \ref{app:analytic}). The velocities in Fig. \ref{parperfig} are computed for two simulations from the right panel of Fig. \ref{fg:figura1}: $v_{d,cr}=c$ with $\lambda=1.5\lambda_{max}$ and $v_{d,cr}=c/2$ with $\lambda=3\lambda_{max}$. These wavelengths correspond to the fastest growing waves for the two $v_{d,cr}$. 
The right panel in Fig. \ref{parperfig}, shows the mean magnitude of the {\it transverse} velocity of plasma particles for the same simulations, and the analytic estimate,  $\vec{v}_{tr,an}\approx fv_{A,0}$. Considering that these two cases keep growing exponentially until $t\gamma_{max} \approx 7$ and 12, respectively (see Fig. \ref{fg:figura1}), we see that, in the $v_A \ll v_{d,cr}$ regime, our analytical estimates for both longitudinal and transverse motions are in good agreement with our numerical results.

\section{Multidimensional effects}\label{sec:multidimensional}

So far we have studied the properties of the CRCD waves assuming an ideal one-dimensional geometry and constant CR current. In this section, we relax the first of these conditions and use two- and three-dimensional PIC simulations to study the CRCD instability, still keeping $\vec{J}_{cr}$ constant. We identify two main differences with respect to the one-dimensional case. The first has to do with the possibility of plasma filamentation that happens before the CRCD instability sets in, as suggested by previous works \citep{NiemiecEtAl08}. We will see below that this filamentation does not occur if the plasma is sufficiently magneitzed, $v_{d,cr}(n_{cr}/n_i) \ll v_{A,0}$. The second multidimensional effect is the interference between CRCD waves generated in different regions of space. Since typically the instability starts from random noise, different regions will give rise to CRCD waves which in general are out of phase with each other. During the non-linear stage, this non-coherence makes the transverse plasma motions from adjacent regions interfere with each other, giving rise to density fluctuations and turbulence in the plasma.

\subsection{Magnetization requirement}\label{sec:magnetization}
\noindent Motivated by previous PIC studies by \cite{NiemiecEtAl08}, we studied the possibility of an initial plasma filamentation that could suppress the formation of the CRCD instability. We ran a series of high space resolution ($c/\omega_{p,e} = 10$ $\Delta$) two-dimensional simulations whose numerical parameters and results are described in Table \ref{table:fil2}. All our two-dimensional simulations are set up in the $x-y$ plane, with $\vec{J}_{cr}$ and $\vec{B}_0$ parallel to the $\hat{x}$ axis. Also, as in some of our one-dimensional simulations, there is a small component of $\vec{B}_0$ pointing along $\hat{z}$, working as a seed for the instability.

\begin{figure*}
\centering
\includegraphics[width=13cm]{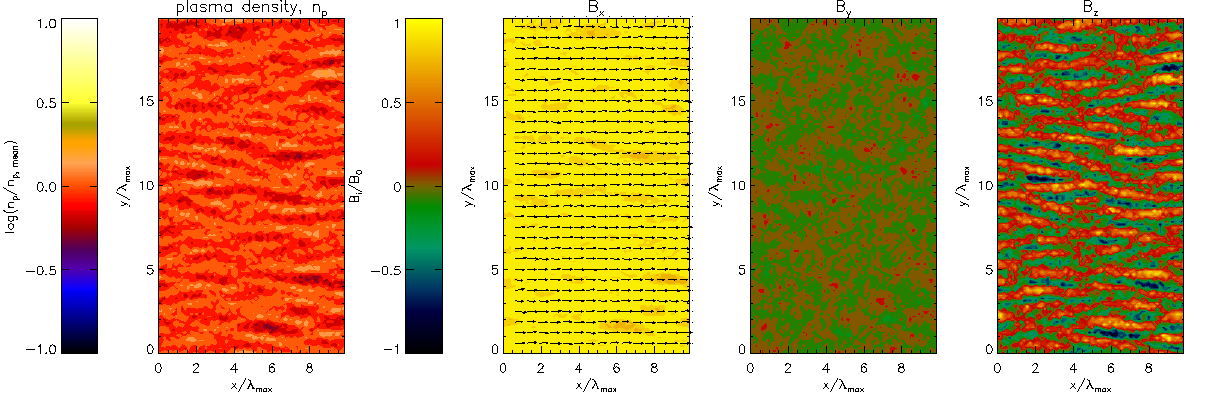}\\
\includegraphics[width=13cm]{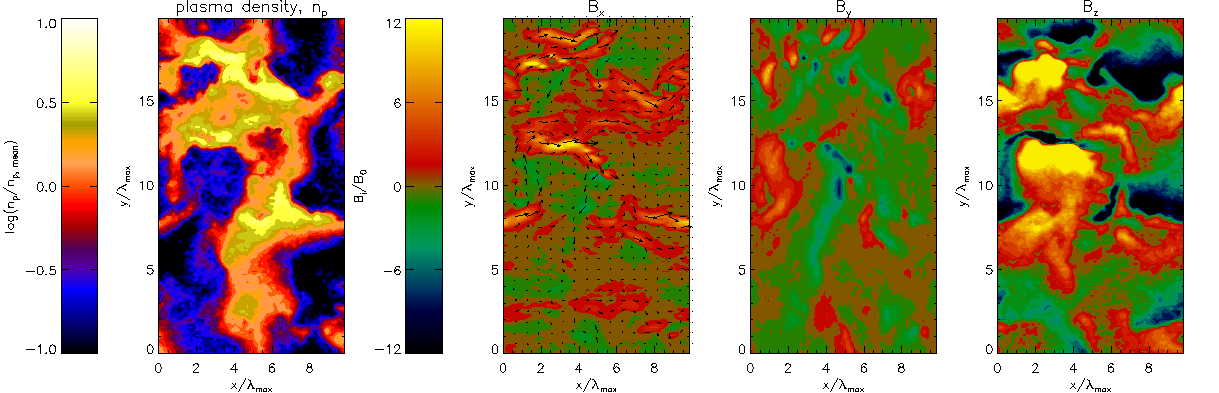}
\caption{Plasma density, $n_p$, and the three components of the magnetic field: $B_x$, $B_y$, and $B_z$ for the two-dimensional simulation M5 (see Table \ref{table:fil2}) at two different times: $t\gamma_{max} = 3$ (top panels) and $11$ (bottom panels), for a box size of $10\lambda_{max} \times 20\lambda_{max}$, where $\gamma_{max}$ and $\lambda_{max}$ are growth rate and wavelength of the fastest growing mode. The arrows in the $B_x$ panels represent the direction of the magnetic field projected on the $x-y$ plane. In this simulation $v_{d,cr}(n_{cr}/n_i) = 4 v_{A,0}$. The Weibel-like plasma filamentation on scales of $\sim10$ electron skin depths can be seen in the plasma density and $B_z$ plots of the top panels. The bottom panels show how the density fluctuations grow both in size and amplitude, suppressing the growth of the CRCD waves.} \label{fg:filamentationc}
\end{figure*}

\begin{figure*}
\centering
\includegraphics[width=13cm]{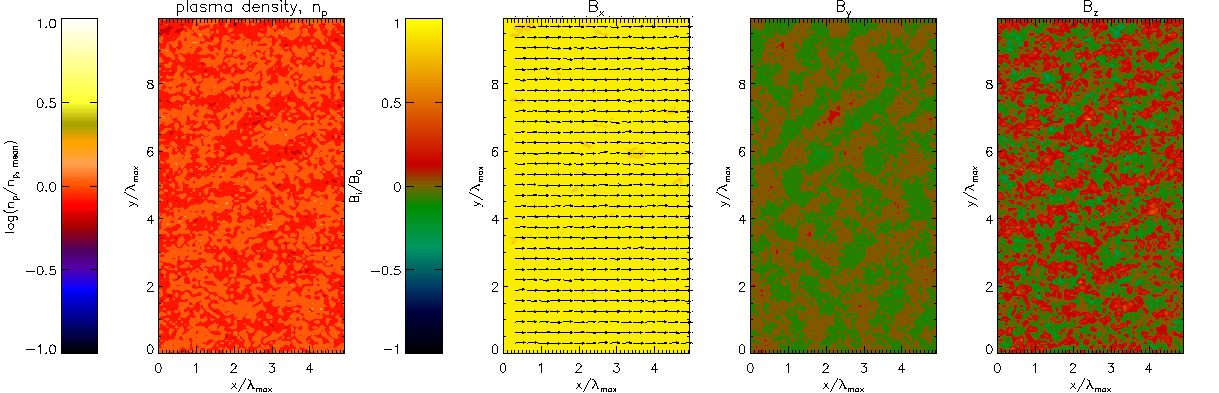}\\
\includegraphics[width=13cm]{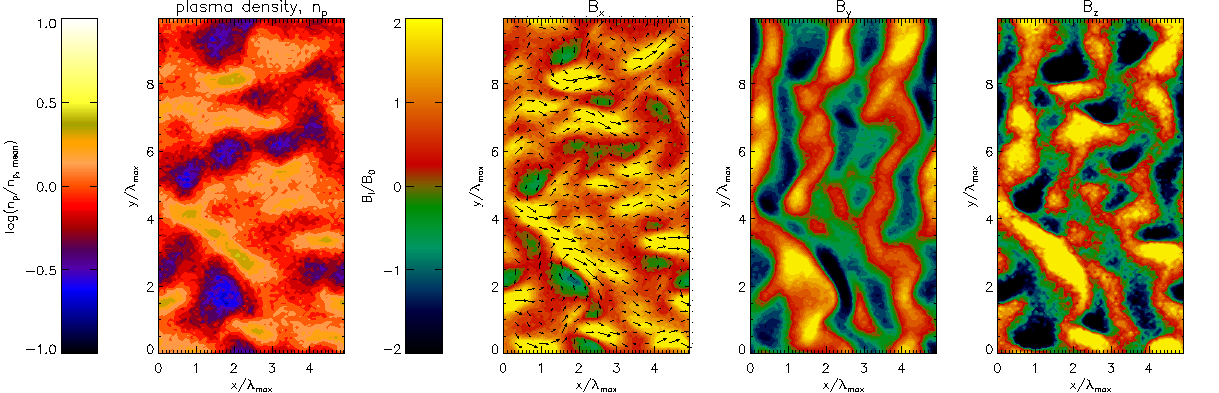}
\caption{Same as in Fig. \ref{fg:filamentationc}, but for simulation M6 of Table \ref{table:fil2}. The only difference with run M5 is that the density of CRs is reduced by half, so $v_{d,cr}(n_{cr}/n_i) = 2 v_{A,0}$ (which implies an increase in $\lambda_{max}$ by a factor of 2). In this case the CRCD waves and the Weibel-like filamentation coexist. We call this situation ``transitional" and represent it by the letter ``T" in Table \ref{table:fil2}.} \label{fg:filamentationb}
\end{figure*}

\begin{figure*}
\centering
\includegraphics[width=13cm]{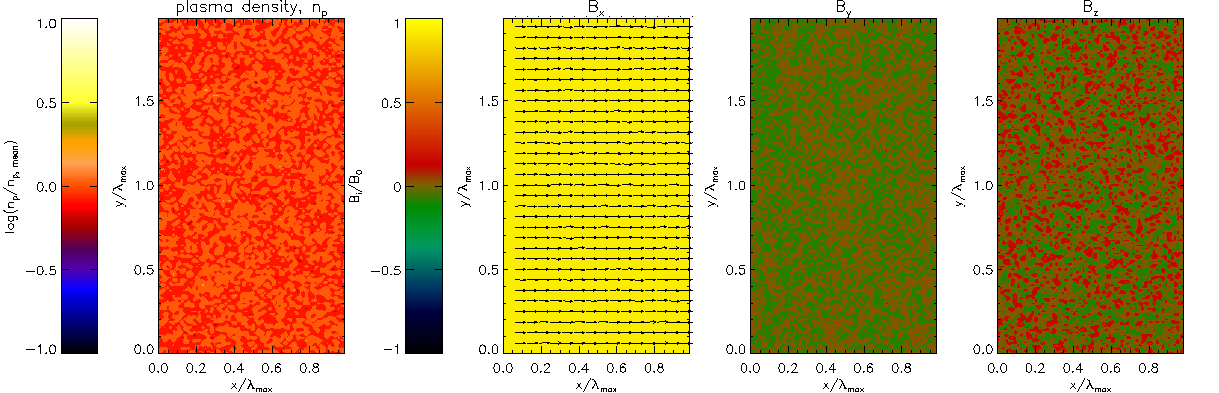}\\
\includegraphics[width=13cm]{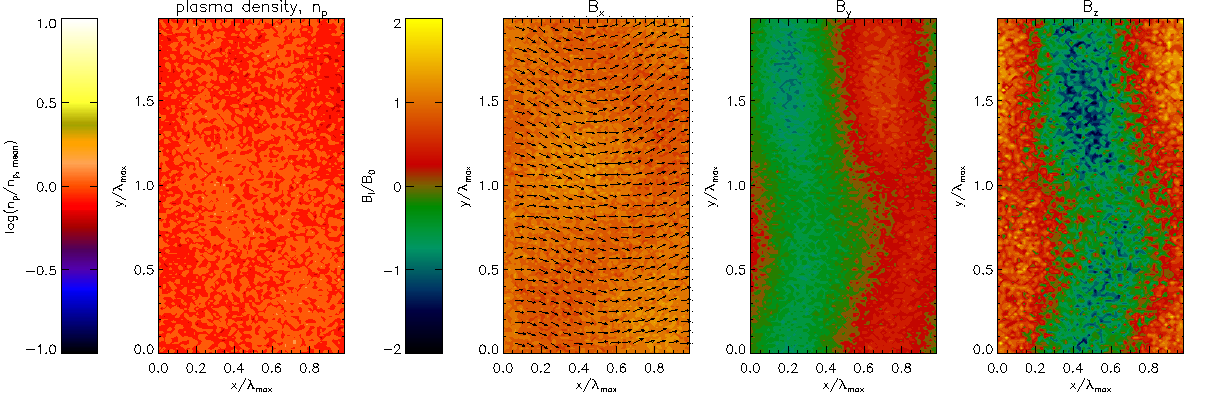}
\caption{Same as in Fig. \ref{fg:filamentationc}, but for the simulation M7 of Table \ref{table:fil2}. The only difference with run M5 is that the density of CRs is reduced by a factor of 10, so $v_{d,cr}(n_{cr}/n_i) = 0.4 v_{A,0}$. We see no Weibel-like filamentation, but a clear CRCD wave of wavelength $\lambda_{max}$.} \label{fg:filamentationa}
\end{figure*}

We identified three regimes, represented in Figs. \ref{fg:filamentationc}, \ref{fg:filamentationb} and \ref{fg:filamentationa}. Fig. \ref{fg:filamentationc} shows the plasma density and three components of the magnetic field for a simulation with $v_{d,cr}(n_{cr}/n_i)/ v_{A,0} = 4$ (run M5 in Table \ref{table:fil2}) at two times $t\gamma_{max} = 3$ and 11. Even though some CRCD field is observed, especially in $B_y$, the dominant instability corresponds to a transverse filamentation that appears initially on the scale of $\sim10$ times the electron skin depth. As time goes on, the filaments merge, creating prominent holes in the plasma that preclude the growth of the instability. Fig. \ref{fg:filamentationa}, on the other hand, shows the same quantities for a simulation where the relative number of CRs was decreased by a factor of 10 (run M7), implying that $v_{d,cr}(n_{cr}/n_i)/v_{A,0} = 0.4$. In this case, a CRCD wave of the size of the box does form (we have chosen the $x$-size of the box to be $\sim \lambda_{max}$). Finally, Fig. \ref{fg:filamentationb} shows the case $v_{d,cr}(n_{cr}/n_i)/v_{A,0} = 2$ (run M6), in which both the CRCD instability and the initial filamentation coexist (we call these cases ``transitional" and indicate them with the letter ``T" in Table \ref{table:fil2}). These three examples indicate that the CRCD instability will develop as long as $v_{d,cr}(n_{cr}/n_i) \ll v_{A,0}$, which is equivalent to having a well magnetized plasma in the sense that $\gamma_{max} \ll \omega_{c,i}$ (see Appendix \ref{app:analytic}). As shown in Table \ref{table:fil2}, we tested the dependence of this criterion on both the magnetization of the plasma (using $\omega_{p,e}/\omega_{c,e} = 3.15$ and 31.5) and the mass ratio, $m_i/m_e$ (using $m_i/m_e = 10$ and 100). We see no difference in our results except that the runs with $m_i/m_e=100$ require a slightly higher value of the ratio $v_{d,cr}(n_{cr}/n_i)/ v_{A,0}$ (a factor of 2 larger) for the transverse filamentation to dominate, but the qualitative criterion remains the same. Also, varying $m_i/m_e$ allows us to determine the physical length scale, $\lambda_{fil}$, at which the transverse filaments appear. This scale shows no dependence on $m_i/m_e$ and corresponds to $\sim10$ times the electron skin depth. Another $m_i/m_e = 100$ simulation was run with zero initial magnetic field (M8), showing that the filaments appear at practically the same scale as in the finite magnetic field case, suggesting a similarity between this filamentation and the Weibel instability. Finally, these results were also tested for a non-relativistic case, $v_{d,cr} = c/2$, obtaining the same conclusions.

Thus, the CRCD instability will grow if the condition $v_{d,cr}(n_{cr}/n_i) \ll v_{A,0}$ (or, equivalently, $\gamma_{max} \ll \omega_{c,i}$) is satisfied. For comparison, the smallest $v_{d,cr}(n_{cr}/n_i)/v_{A,0}$ factor used by \cite{NiemiecEtAl08} is 1.31, which is close to the regime where the transverse filaments appear. This fact would explain their plasma filamentation, which may have suppressed the appearance of the CRCD instability.


\begin{table*}
\begin{center}
\caption{Numerical parameters and results for the ``M" simulations.}
\label{table:fil2}
\begin{tabular}{|l|l|l|l|l|l|l|l|l|l|l|l|l|}%|c|c||c|c|}
\hline
Run & $m_i/m_e$ & $n_{cr}/n_i$ & $v_{A,0}/v_{d,cr}$ & $v_{d,cr}$& $\omega_{p,e}/\omega_{c,e}$ &$\gamma_{max}^{-1}$ & $\omega_{c,i}/\gamma_{max}$ &  $\gamma/\gamma_{max}$ & $L_x\times L_y$ & $\lambda_{max}$ & $\lambda/\lambda_{max}$ & $\lambda_{fil}/\lambda_{s,e}$ \\
& & & & & & ($\Delta_t$) & & & ($\Delta^2$) & ($\Delta$) & & \\
\hline
\hline
M1 & 10 & 0.4 & 1/10 & c & 3.15 & 352 & 0.5 &- & 1024$\times$2048 & 99 & - & 8.5\\
\hline
M2 & 10 & 0.2 & 1/10 & c & 3.15 & 703 & 1 &T & 1024$\times$2048 & 199 & T & T\\
\hline
M3 & 10 & 0.04 & 1/10 & c & 3.15 & 3,516 & 5 &0.82 & 1024$\times$2048 & 994 & 1.03 & -\\
\hline
\hline
M4 & 10 & 0.2 & 1/100 & c & 31.5 & 703 & 0.1 &- & 1024$\times$2048 & 20 & - & 9.8\\
\hline
M5 & 10 & 0.04 & 1/100 & c & 31.5 & 3,516 & 0.5 &- & 1024$\times$2048 & 99 & - & 8.5\\
\hline
M6 & 10 & 0.02 & 1/100 & c & 31.5 & 7,031 & 1 &T & 1024$\times$2048 & 199 & T & T\\
\hline
M7 & 10 & 0.004 & 1/100 & c & 31.5 & 35,156 & 5 &0.43 & 1024$\times$2048 & 994 & 1.03 & -\\
\hline
\hline
M8 & 100 & 0.1264 & 0 & c & $\infty$ & - & 0 &- & 3500$\times$1024 & - & - & 13\\
\hline
M9 & 100 & 0.1264 & 1/63.24 & c & 6.31 & 3,516 & 0.25 &- & 3500$\times$1024 & 157 & - & 9.3\\
\hline
M10 & 100 & 0.0632 & 1/63.24 & c & 6.31 & 7,031 & 0.5 &T & 3500$\times$1024 & 314 & T & T\\
\hline
M11 & 100 & 0.0316 & 1/63.24 & c & 6.31 & 14,063 & 1 &0.43 & 3500$\times$1024 & 629 & 0.7 & -\\
\hline
\hline
M12 & 10 & 0.253 & 1/15.81 & c/2 & 10 & 1,111 & 0.5 &- & 2048$\times$2048 & 95 & - & 8.5\\
\hline
M13 & 10 & 0.1265 & 1/15.81 & c/2 & 10 & 2,222 & 1 &T & 2048$\times$2048 & 190 & T & T\\
\hline
M14 & 10 & 0.0632 & 1/15.81 & c/2 & 10 & 4,444 & 2 &0.69 & 2048$\times$2048 & 379 & 0.8 & -\\
\hline
\end{tabular}
\tablecomments{We list the ion-electron mass ratio $m_i/m_e$, the CR-ion number density ratio $n_{cr}/n_i$, the ratio between the initial Alfv\'en velocity of the plasma and the drift velocity of the CRs $v_{A,0}/v_{d,cr}$, the CR drift velocity $v_{d,cr}$, the ratio between the plasma and the cyclotron frequencies of electrons $\omega_{p,e}/\omega_{c,e}$, the inverse of the maximum theoretical growth rate $\gamma_{max}^{-1}$ in units of the time step $\Delta_t$, the ratio between the cyclotron frequency of ions and $\gamma_{max}$, the measured growth rate $\gamma$ divided by $\gamma_{max}$, the dimensions of the simulation box $L_x\times L_y$ in units of of the grid cell size $\Delta$, the theoretical wavelength of maximum growth $\lambda_{max}$ in units of $\Delta$, the measured wavelength of maximum growth $\lambda$ divided by $\lambda_{max}$, and the size of the transverse Weibel-like filaments $\lambda_{fil}$ in units of the electron skin depth $\lambda_{s,e}$ ($=c/\omega_{p,e}$). The symbol ``-" and the letter ``T" indicate the cases with no CRCD field growth and the ones with a combination of CRCD growth and Weibel-like filamentation, respectively. In all the runs the number of particles per cell, $N_{ppc}$, is 4.}
\end{center}
\end{table*}

\subsection{Multidimensional Evolution}
\begin{table*}
\begin{center}
\caption{Numerical parameters of the ``I" simulations.}
\label{table:interference}
\begin{tabular}{|l|l|l|l|l|l|l|l|l|l|l|}%|c|c||c|c|}
\hline
Run & $L_x\times L_y(\times L_z)/\Delta^3$  & $\lambda_{s,e}/\Delta$ & $\lambda_{max}/\Delta$  & $m_i/m_e$ & $v_{A,0}/v_{d,cr}$ & $v_{d,cr}$ & $\omega_{p,e}/\omega_{c,e}$ & $\gamma_{max}^{-1}/\Delta_t$ & $\omega_{c,i}/\gamma_{max}$ & $N_{ppc}$\\
\hline
\hline
I1 & $512^3$ & 0.5 & 32 & 10 & 1/40 & c & 12.65 & 2083 & 3.22 &  4\\ 
\hline
I2 & $512^3$ & 0.5 & 32 & 10 & 1/20 & c/2 & 12.65 & 2083 & 3.22 &  4\\
\hline
I3 & $512^3$ & 0.5 & 32 & 10 & 1/10 & c & 3.16 & 521 & 3.22 &  4\\
\hline
\hline
I4 & 2048$\times$4096 & 1.25 & 124 & 10 & 1/31.62 & c & 10 & 2774 & 5 & 4\\
\hline
I5 & 1024$\times$2048 & 2.5 & 149 & 10 & 1/10 & c & 3.17 & 949 & 3 &  4\\
\hline
I6 & 1024$\times$2048 & 2.5 & 149 & 10 & 1/50 & c & 15.8 & 4743 & 3 & 4\\
\hline
\end{tabular}
\tablecomments{We list the dimensions of the simulation box $L_x\times L_y (\times L_z)$ , the electron skin depth $\lambda_{s,e}$ ($=c/\omega_{p,e}$), and the theoretical wavelength of maximum growth $\lambda_{max}$ in units of $\Delta$. Also, we list the ion-electron mass ratio $m_i/m_e$, the ratio between the initial Alfv\'en velocity of the plasma and the drift velocity of the CRs $v_{A,0}/v_{d,cr}$, the CRs drift velocity $v_{d,cr}$, the ratio between the plasma and the cyclotron frequencies of electrons $\omega_{p,e}/\omega_{c,e}$, the inverse of the maximum theoretical growth rate $\gamma_{max}^{-1}$ in units of the time step $\Delta_t$, the ratio between the cyclotron frequency of ions and $\gamma_{max}$, and the number of particle per cell $N_{ppc}$.}
\end{center}
\end{table*}
We studied CRCD instability with a series of two- and three-dimensional simulations whose numerical parameters are summarized in Table \ref{table:interference}. As we will see below, when multidimensional effects are considered, the dominant wavelength of the instability, $\lambda_d$, is initially equal to $\lambda_{max}$ but then rapidly grows as the field is amplified. This can make $\lambda_d$ equal to the size of the box $L$ before the instability reaches saturation, which can make sufficiently large three-dimensional simulations challenging. We discuss here the results of our three-dimensional runs, and check them in Appendix \ref{sec:2d} with large two-dimensional simulations, for which $\lambda_d$ is always significantly smaller than the size of the box. 

\subsubsection{Three-dimensional Simulations}\label{sec:threed}

%Although in the three-dimensional case it is difficult to follow the non-linear amplification of the field without making $\lambda_d$ larger than $L$, we use three-dimensional simulations to make a more realistic analysis of the spacial structure of the CRCD instability. Below 
In this section we present the results of three three-dimensional simulations that test saturation in the non-relativistic and relativistic regimes, and the dependence of the amplification on the initial magnetic field and the CR drift velocity. Two of the simulations have the same $v_{A,0}/c = 1/40$, but $v_{d,cr} = c$ and $c/2$ (runs I1 and I2 in Table \ref{table:interference}). The third simulation has $v_{A,0}/c = 1/10$ and $v_{d,cr} = c$ (run I3 in Table \ref{table:interference}). As in all our simulations so far, $\vec{J}_{cr}$ and $\vec{B}_0$ point along $\hat{x}$ (apart from a small component of the magnetic field along $\hat{z}$ of magnitude $B_0/10$), and the back-reaction on the CRs is not included. The rest of the numerical parameters are specified in Table \ref{table:interference}.

\begin{figure*}
\centering\includegraphics[width=13cm]{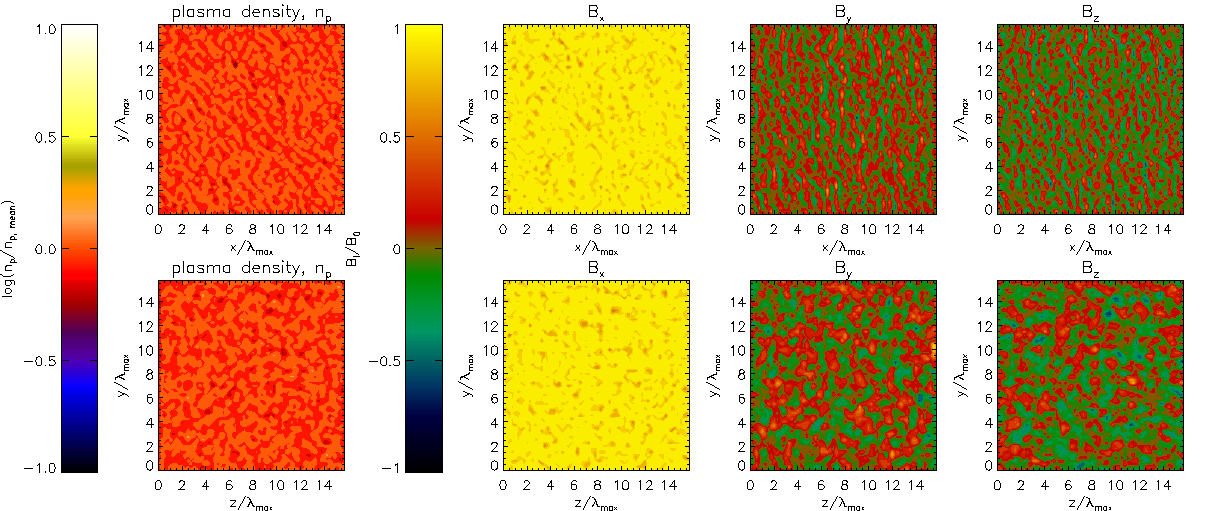}
\caption{Plasma density, $n_p$, normalized using the mean plasma density, $n_{p,mean}$, and the three components of the magnetic field: $B_x$, $B_y$, and $B_z$, normalized in terms of $B_0$, for the three-dimensional simulation I1 described in Table \ref{table:interference}. The top and bottom panels correspond to longitudinal ($z/\lambda_{max} = 8$) and transverse ($x/\lambda_{max} = 8$) slices of the simulation box, respectively. At this moment ($t\gamma_{max}=4$), the CRCD waves are still in the linear regime, growing exponentially at a rate $\sim \gamma_{max}$ (see Fig. \ref{fg:departure3d}) and at a preferred wavelength $\sim \lambda_{max}$.} \label{fg:cortes1}
\end{figure*}
\begin{figure*}
\centering\includegraphics[width=13cm]{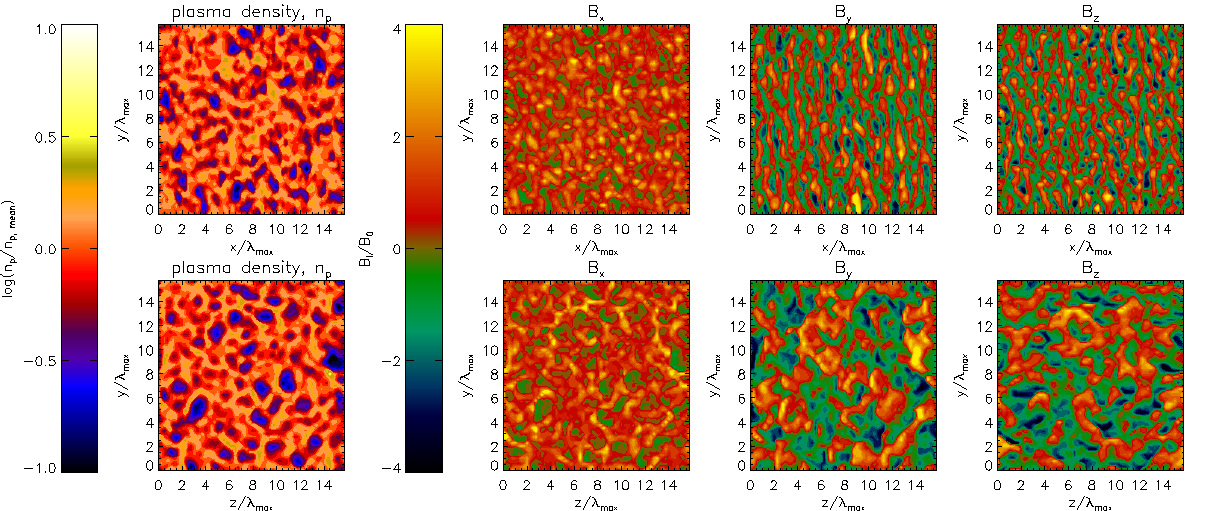}
\caption{Same as in Fig. \ref{fg:cortes1} but for $t\gamma_{max}=7$ (when $B_{tr} \sim B_0$). Prominent holes are starting to form in the plasma ($\Delta n_p/n_p \sim 10$) on scales of $\sim \lambda_{max}$. Also the CRCD waves are starting to get distorted by the turbulent motions in the plasma. Also, the transverse slices show how the transverse spacial correlation of the waves have increased compared to the $t\gamma_{max}=4$ case (depicted in Fig. \ref{fg:cortes1}).}\label{fg:cortes2}
\end{figure*}
\begin{figure*}
\centering\includegraphics[width=13cm]{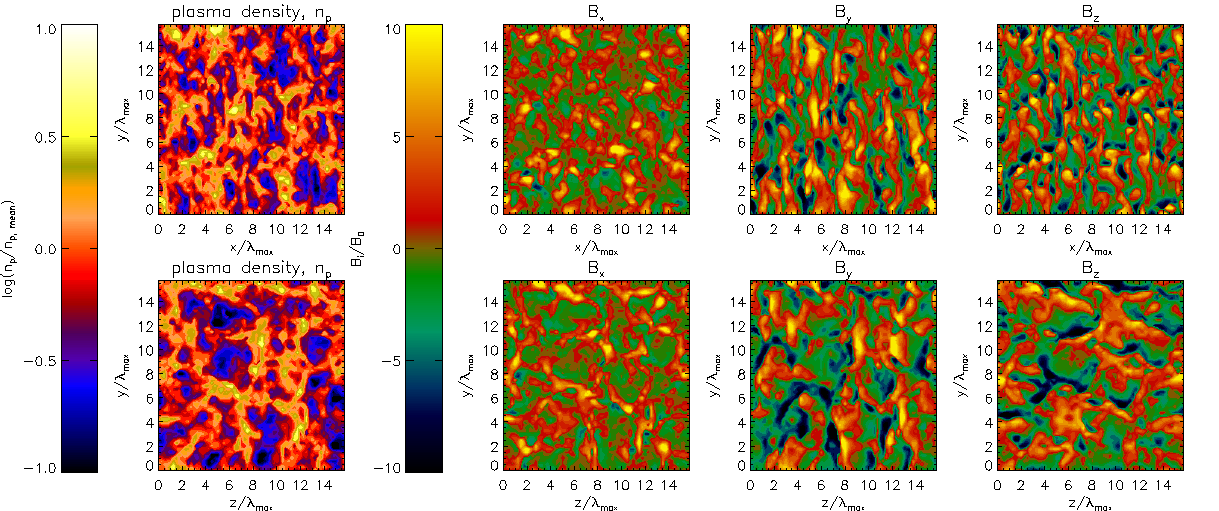}
\caption{Same as in Fig. \ref{fg:cortes1} but for $t\gamma_{max}=9$ ($B_{tr} \gg B_0$, see black line in Fig. \ref{fg:departure3d}). The qualitative features of the turbulence are the same than in Fig. \ref{fg:cortes2}, except for an increase in the size of the plasma holes and dominant CRCD wavelength.} \label{fg:cortes3}
\end{figure*}
\begin{figure*}
\centering\includegraphics[width=13cm]{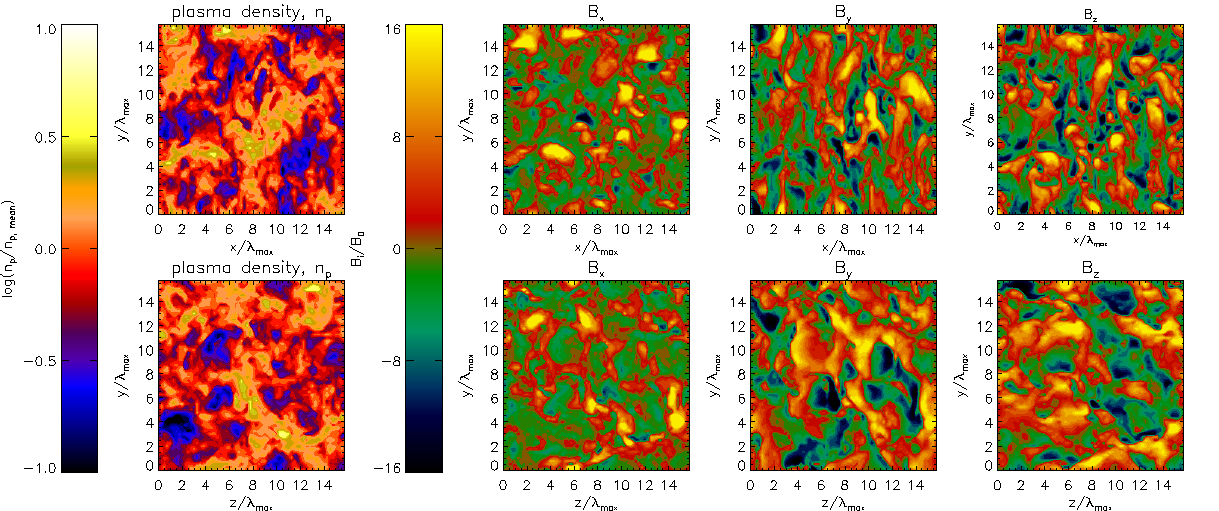}
\caption{Same as in Fig. \ref{fg:cortes1} but for $t\gamma_{max}=11$. Here the size of the magnetic fluctuations have become close to the size of the simulation box. The turbulence looks similar to Fig. \ref{fg:cortes3}.} \label{fg:cortes4}
\end{figure*}
\begin{figure}
\centering\includegraphics[width=8cm]{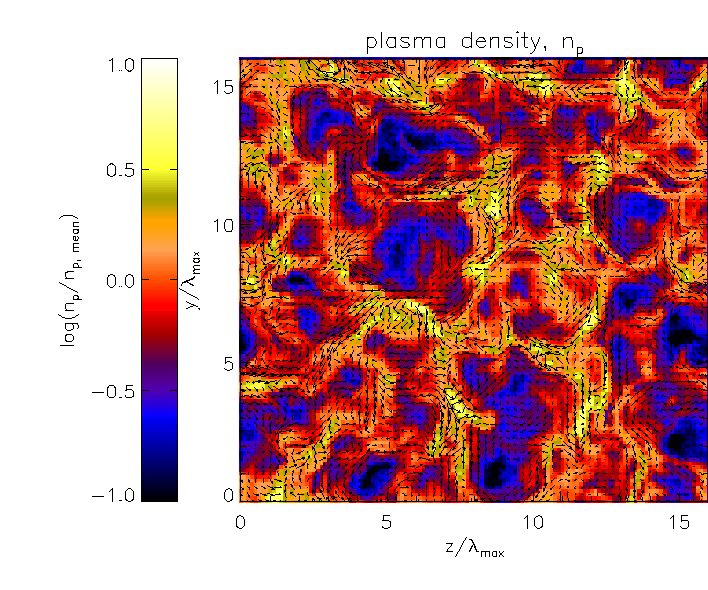}
\caption{The density plot presented in Fig. \ref{fg:cortes3}, but with overplotted arrows showing the magnetic field projection on the $z-y$ plane. The clock-wise orientation of the magnetic field lines around the plasma holes shows the presence of CRCD waves driving the turbulence.} \label{fg:zoomcorte}
\end{figure}

\begin{figure}
\centering\includegraphics[width=7.75cm]{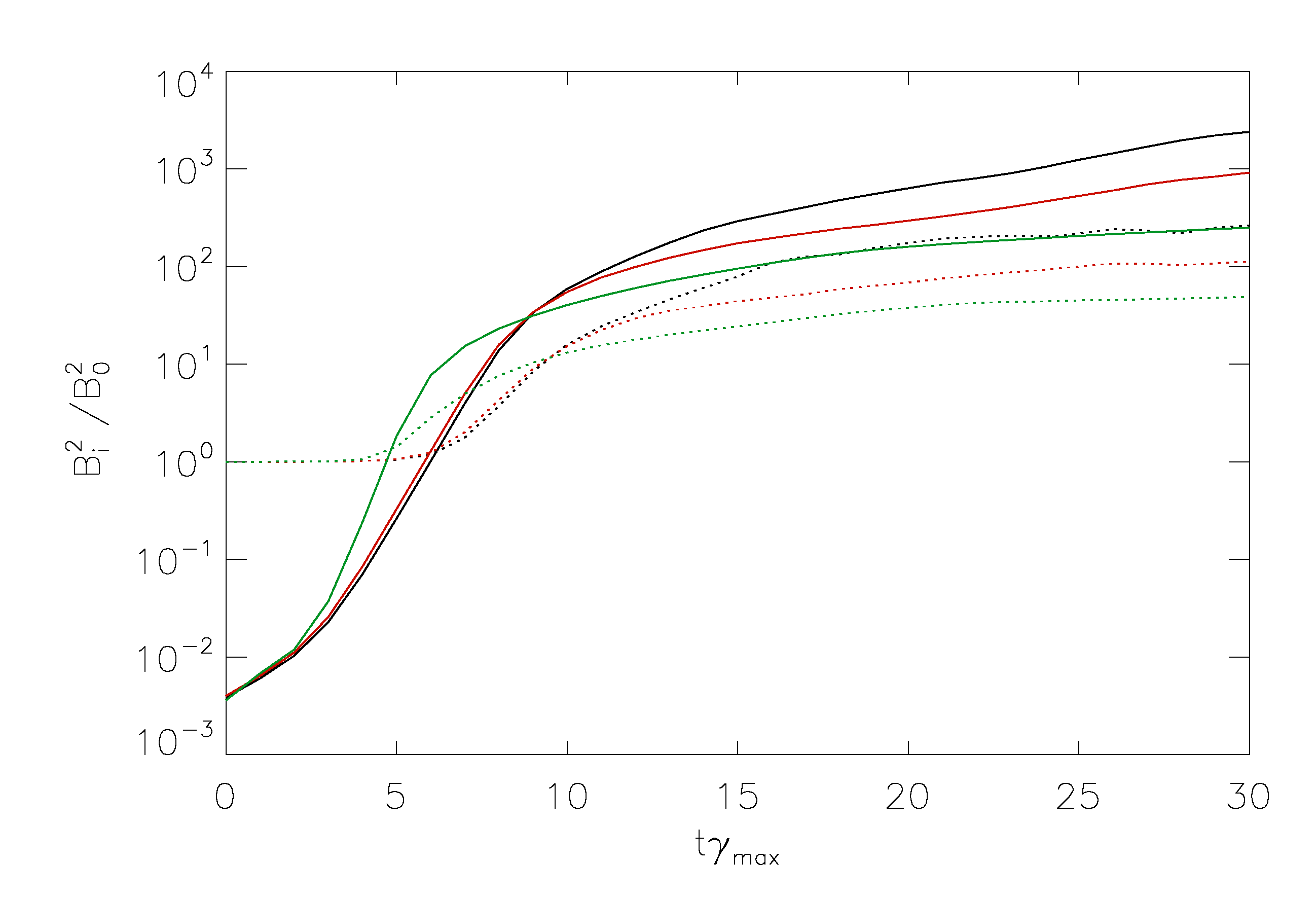}
\caption{The transverse (solid) and longitudinal (dotted) components of the magnetic energy as a function of time for three-dimensional simulations I5 (black), I6 (red), and I7 (green) of Table \ref{table:interference}, for which ($v_{A,0}/v_{d,cr}, v_{d,cr}/c$) = (1/40,1), (1/20,0.5), and (1/10,1), respectively. Time is normalized in terms of the $\gamma_{max}$ of each simulation. In all the runs, the departure from exponential growth occurs  after $B_{tr} \sim B_0$, but saturation happens at $v_{A} \sim v_{d,cr}$.} \label{fg:departure3d}
\end{figure}

The evolution of the plasma density and the three components of the magnetic field for simulation I1 can be seen in Figs. \ref{fg:cortes1}, \ref{fg:cortes2}, \ref{fg:cortes3}, and \ref{fg:cortes4}. These figures show two slices of the simulation box. One is longitudinal and corresponds to the plane $z/\lambda_{max} = 8$ (top panels), and the other is transverse and corresponds to $x/\lambda_{max} = 8$ (bottom panels), where $\lambda_{max} = L/16$. The magnetic energy evolution for the same run is depicted in Fig. \ref{fg:departure3d}.

Fig. \ref{fg:cortes1} shows the early moments of the instability ($t = 4\gamma_{max}^{-1}$). The longitudinal slice shows how the CRCD waves form independently in different regions of the box. This is also seen in the transverse slice, which shows how the phases of the waves differ between different points of the plane $x/\lambda_{max} = 8$.

Fig. \ref{fg:cortes2} shows the beginning of the non-linear regime ($f \sim 1$), which corresponds to $t\gamma_{max}=7$. In this case, the phases of the waves are transversely more correlated compared to Fig. \ref{fg:cortes1}, as can be seen in the plots of the transverse slice for $B_y$ and $B_z$. This increased spatial correlation indicates that, until this moment, the adjacent waves were merging without significantly interfering with each other. Also, Fig. \ref{fg:cortes2} shows the appearance of prominent density fluctuations ($\Delta n_p/n_p \sim 10$, where $n_p$ is the plasma density). As we saw in \S \ref{sec:thewaves}, when the CRCD waves are in the exponential growth regime ($v_A \ll v_{d,cr}$), the background plasma will move transversely at $v_{tr,an} \approx v_{A,0}f$. So, when the instability gets non-linear, the transverse velocity of the plasma becomes close to $v_A$. In a low temperature regime, this velocity corresponds to the magnetosonic sound speed of the plasma. Thus, as soon as $B_{tr} \sim B_0$, the transverse motions will produce moderate shocks, giving rise to significant density fluctuations in the plasma. At this point the transverse plasma motions develop into isotropic turbulence with velocities of the order of $v_{A,0}f$.
What happens after the density fluctuations appear can be seen in Fig. \ref{fg:cortes3} ($t = 9\gamma_{max}^{-1}$). The longitudinal slice shows how the magnetic fluctuations get distorted and increase rapidly in size. As already mentioned in \S \ref{sec:nonrel}, even in one-dimensional geometry the instability is expected to evolve into wavelengths longer than $\lambda_{max}$. However, in a multidimensional set-up, the evolution into magnetic fluctuations of larger size gets accelerated after the appearance of the density fluctuations and turbulence in the plasma. We will quantify this migration in \S \ref{sec:migration}. The transverse slice of Fig. \ref{fg:cortes3} shows how the underdense regions (or holes) have merged and increased their size with respect to Fig. \ref{fg:cortes2}. It is also interesting to see from Fig. \ref{fg:zoomcorte} how the holes are separated by ``plasma walls" through which the transverse magnetic field reverses direction. We see that the magnetic field has a clockwise orientation around the holes, which is consistent with the presence of right-handed waves producing the expansion of the holes.

Finally, Fig. \ref{fg:cortes4} ($t = 11\gamma_{max}^{-1}$) shows essentially no difference with respect to Fig. \ref{fg:cortes3} besides the growth of the size of both the magnetic fluctuations and the plasma holes, which at this point are close to $L$.

Fig. \ref{fg:departure3d} shows the magnetic energy evolution for the three-dimensional simulations. We can see that, in the three cases, the departure from the exponential growth occurs shortly after the wave becomes non-linear (which coincides with the generation of significant density fluctuations and turbulent motions in the plasma). We also see that the final saturation satisfies the $v_A \sim v_{d,cr}$ condition, which suggests that, when multidimensional effects are included, the intrinsic saturation of the CRCD instability is still given by the $v_A \sim v_{d,cr}$ criterion. Unfortunately, in our three-dimensional simulations, this saturation happens when the dominant sizes of the holes and magnetic fluctuations have already become close to $\sim L$. Saturation still happens at $v_A \sim v_{d,cr}$ because, when $\lambda_d = L$, the three-dimensional simulations behave more like the one-dimensional simulations presented in \S \ref{sec:rel}, in the sense that there is only one dominant mode that saturates at $v_A \sim v_{d,cr}$. After $\lambda_d = L$, the density fluctuations almost disappear and the turbulent motions transform into more coherent transverse plasma motions. In any case, the $v_A \sim v_{d,cr}$ saturation criterion is confirmed by two-dimensional simulations presented in Appendix \ref{sec:2d} for which $\lambda_d$ is always smaller than $L$.

Fig. \ref{fg:departure3d} also shows that, in the three-dimensional simulations, the magnitude of the magnetic component along $\hat{x}$ is comparable to the transverse one, suggesting a rather isotropic orientation of the CRCD field.

\subsubsection{Migration into longer wavelengths}\label{sec:migration}
\begin{figure}
\centering\includegraphics[width=8cm]{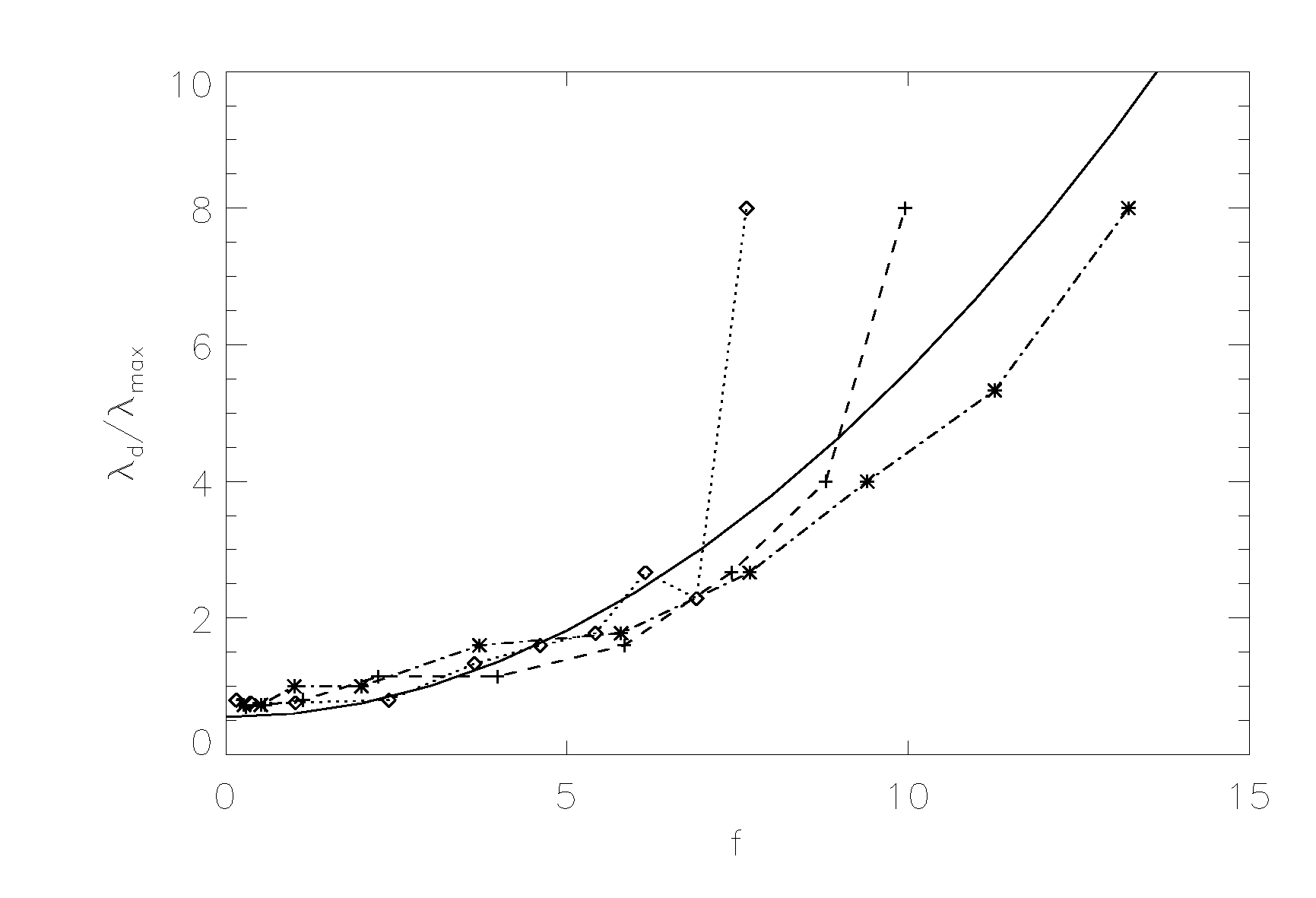}
\caption{Evolution of the dominant wavelength, $\lambda_d$, as a function of the amplification factor, $f$, for three-dimensional simulations I1 (dot-dashed), I2 (dashed), and I3 (dotted). For comparison, our semi-analytical formula, $\lambda_d = \lambda_{max}((f/3)^2+1)/2$, is shown as solid line.}\label{fg:wavelength}
\end{figure}
Even though migration to longer wavelengths is already observed in one-dimensional simulations, it becomes faster when multidimensional effects are considered. In this section we propose a semi-analytic model that quantifies this migration in terms of the amplification factor of the field, $f$. As we saw in \S \ref{sec:threed}, the motions associated with the turbulence tend to distort the CRCD waves, producing a damping of the shortest wavelength modes. Thus, the dominant wavelength, $\lambda_d$, will correspond to the fastest growing mode that can be amplified without being strongly affected by the turbulence. Considering that a CRCD wave of wavelength $\lambda$ grows in a time scale comparable to the inverse of its growth rate $\gamma^{-1}(\lambda)$ from Equation (\ref{eq:dispersion}), and that the turbulence will kill it in a time scale comparable to $\lambda/v_{turb}$, where $v_{turb}$ is the typical turbulent velocity, then $\lambda_d$ will be such that $\gamma^{-1}(\lambda_d) \sim \lambda/v_{turb}$. Since the turbulence is due to the transverse plasma motions produced by non-coherent CRCD waves, then $v_{turb}$ must be comparable to the transverse velocity of the waves, which we already determined to be $fv_{A,0}$. So, $\lambda_d$ will be such that $\gamma(\lambda_d) \approx \beta v_{A,0}f/\lambda_d$, where $\beta$ is an unknown constant that quantifies the relative importance of the two time scales. If we get $\gamma(\lambda_d)$ from the real part of Equation (\ref{eq:dispersion}), we can obtain $\beta$ by fitting the evolution of $\lambda_d$ in our three-dimensional simulations, obtaining $\beta \approx 2$. This way we find $\lambda_d$ as a function of $f$,
\begin{equation}
\lambda_d \approx \lambda_{max}[(f/3)^2 + 1]/2, 
\label{eq:wavelength}
\end{equation}
which is intended to be valid after the turbulence becomes significant ($f \gtrsim 3$). In Fig. \ref{fg:wavelength} we show a comparison between this formula and the evolution of $\lambda_d$ as a function of $f$ for our three-dimensional simulations. We computed $\lambda_d$ by performing Fourier transforms of $B_y$ and $B_z$ along lines of constant $y$ and $z$ coordinates, and then finding the mean wavelength of the peak of the Fourier transform. As the dominant wavelengths approach $L$ ($=16\lambda_{max}$), the determination of $\lambda_d$ becomes quite noisy. Due to this reason, we have plotted our simulation results only until $\lambda_d = L/2$. We see from Fig. \ref{fg:wavelength} that Equation (\ref{eq:wavelength}) appears to provide an acceptable fit for the evolution of $\lambda_d$. Our result shows a growth of $\lambda_d$ substantially faster than the direct proportionality between $\lambda_d$ and $f$ suggested by \cite{Bell04}. We will see below that, when the back-reaction on the CRs is considered, this difference has important implications to the saturation of the CRCD instability.

\section{Back-reaction on cosmic rays}\label{freecrs}

\begin{figure}
\centering\includegraphics[width=8cm]{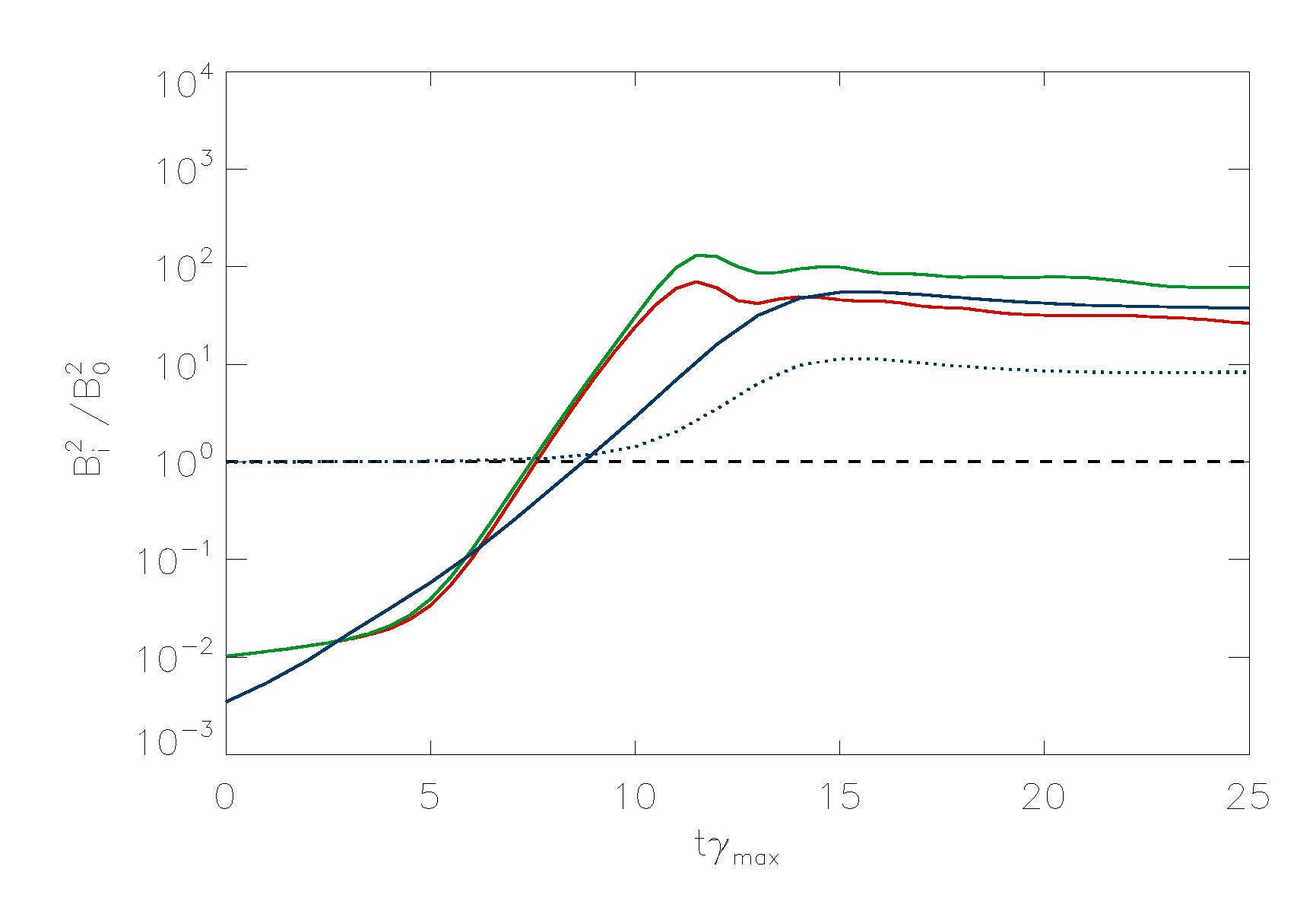}
\caption{$B_{tr}^2$ and $B_x^2$ normalized in terms of $B_{0}^2$, are plotted as a function of time (normalized using $\gamma_{max}$) for one- and three-dimensional runs in order to study the effect on the magnetic energy evolution due to the back-reaction on the CR. In all cases CRs are monoenergetic and have a semi-isotropic momentum distribution such that $v_{d,cr} = c/2$. The two one-dimensional simulations have numerical parameters: $n_{cr}/n_i = 0.005$, $v_{d,cr}/c = 0.5$, $v_{A,0}/v_{d,cr} = 1/40$, $\lambda_{max} = 2048$ $\Delta$, $L/\lambda_{max} = 15$, $m_i/m_e = 100$, $\omega_{p,e}/\omega_{c,e} = 8$, $\gamma_{max}^{-1} = 231,786$ $\Delta_t$, $\omega_{c,i}/\gamma_{max} = 10$, and $c/\omega_{p,e} = 3.3$. Their CR Lorentz factor $\Gamma$ is 20 (solid, red line) and 40 (solid, green line), respectively. The three-dimensional simulation, whose transverse magnetic energy is represented by the solid, blue line, has the same parameters as run I2 in Table \ref{table:interference}, but with $\Gamma = 30$. The dotted, blue line represents the longitudinal magnetic energy ($B_x^2/B_0^2$). The dashed, black line shows the constant magnetic energy along $\hat{x}$ for the two one-dimensional simulations.} \label{fg:unlocked}
\end{figure}
In this section we use one- and three-dimensional simulations to study the effect of the dynamic evolution of the CRs on the saturation of the CRCD instability. We will concentrate on the case of a beam of monoenergetic CRs drifting at half the speed of light ($v_{d,cr} = 0.5c$) in the $\hat{x}$ direction, which is parallel to $\vec{B_0}$ (except for a small magnetic component along $\hat{z}$). This drift is obtained by sampling the CR velocities from an isotropic, monoenergetic momentum distribution, but only keeping the velocities in the positive $x$ direction. This choice for the CR momentum distribution has a direct application to the  most energetic CRs that propagate in the upstream medium of SNR shocks (see \S\ref{conclusions}).

The red and green curves in Fig. \ref{fg:unlocked} represent the magnetic energy evolution for two one-dimensional simulations that only differ in their Lorentz factors, $\Gamma$, taken to be 20 and 40 (the rest of the numerical parameters are specified in the caption of Fig. \ref{fg:unlocked}). These simulations saturate at $f \approx 7.9$ and $12.6$, when the Larmor radius of CRs is close to the dominant wavelength of the instability ($R_{L,cr}/\lambda_d = 1.05$ and $1.12$ for two runs). To obtain this result we use that initially $R_{L,cr}/\lambda_{max} = (n_{cr}/n_i)(v_{d,cr}/c)(c/v_{A,0})^2\Gamma/4\pi$, and consider that at saturation the dominant wavelength has grown ($\lambda_d/\lambda_{max} \approx 2.9$ and $3.4$, respectively) and CRs have lost part of their energy (the mean $\Gamma$ of CRs is 18.9 and 37.6, respectively). These results are confirmed by a three-dimensional simulation whose magnetic energy evolution is represented by the blue lines in Fig. \ref{fg:unlocked}. The numerical parameters of this three-dimensional simulation are the same as in run I2 (see Table \ref{table:interference}) but includes the back-reaction on the CRs, whose $\Gamma = 30$. If we consider that at saturation $\lambda_d/\lambda_{max} = 4$ and the mean $\Gamma$ of the CRs is 27.6, we obtain that the CR deflection saturates the instability when $R_{L,cr}/\lambda_d = 1.35$. Note that in this simulation $\lambda_d$ is always a factor of 4 smaller than the size of the box, so the saturation is not affected by box effects. For the semi-isotropic distribution of monoenergetic CRs presented here, we find that the saturation due to CR back-reaction will happen when $R_{L,cr} \approx \lambda_{d}$. Although this result is valid for our particular choice of CR momentum distribution, we expect that in general the saturation of the CRCD instability will be determined either by the intrinsic limit $v_A \sim v_{d,cr}$, or by the strong CR deflection when $R_{L,cr} \sim \lambda_{d}$. As will be discussed in \S \ref{conclusions}, achieving $v_A \sim v_{d,cr}$ requires a very high CR energy density, a condition that is not expected for non-relatistic shocks environments.

Also, our simulations show that, at saturation, many CRs have negative $x$ velocity. This suggests that, besides the field amplification, the CRCD instability can provide an efficient scattering mechanism for CRs upstream of shocks.

\section{Discussion and Conclusions}\label{conclusions}
Using fully kinetic PIC simulations, we confirmed the existence of the CRCD instability predicted by Bell (2004). Combining one-, two-, and three-dimensional simulations with an analytic, kinetic model we studied the non-linear properties of the instability and its possible saturation mechanisms.

In the first part, we studied non-linear CRCD waves under idealized conditions, namely: \emph{i}) ignoring multidimensional effects, and \emph{ii}) assuming a constant CR current without back-reaction on the CRs. We confirm that the CRCD waves can grow exponentially at the wavelengths and rates predicted by the analytic dispersion relation \citep{Bell04, RevilleEtAl06, BlasiEtAl07}. We find that the exponential growth can continue into the very nonlinear regime, until the Alfven velocity in the amplified field is comparable to the CR drift velocity, $v_A \sim v_{d,cr}$. This saturation is due to plasma acceleration along the direction of motion of the CRs, which reduces the CR current observed by the plasma particles. The plasma moves at the velocity $\sim f^2v_{A,0}^2/v_{d,cr}$, where $f$ is the amplification factor of the field ($f \equiv B_{tr}/B_0$). At saturation, when $v_A \approx v_{A,0}f \sim v_{d,cr}$, the plasma moves together with CRs, decreasing the net driving current. The waves also induce transverse plasma motions with velocities $\sim v_{A,0}f$. These motions generate plasma turbulence when multidimensional effects are included.
 
In the second part, we considered more realistic conditions by including the multidimensional effects using two- and three-dimensional simulations with constant $\vec{J}_{cr}$. Our main results are: 

\emph{i)} In the linear regime, if the plasma is well magnetized ($\gamma_{max} \ll \omega_{c,i}$, or, equivalently, $v_{d,cr}(n_{cr}/n_i) \ll v_{A,0}$), the CRCD waves grow at the rate and preferred wavelength close to the ones obtained in the one-dimensional analysis. If this condition is not met, Weibel-like filaments form in the plasma, supressing the appearance of the waves. In this case, the streaming CRs can still amplify the magnetic field to non-linear values, but at a rate significantly lower than that of the CRCD waves \citep{NiemiecEtAl08}. This regime, however, might be relevant to the upstream medium of relativistic shocks in GRBs, where $n_{cr}$ could exceed $n_i$ \citep{CouchEtAl08}.

\emph{ii)} In the non-linear CRCD regime, the transverse plasma motions associated with the instability create significant density fluctuations and turbulence in the plasma. These turbulent motions suppress the growth of the shortest CRCD waves, producing a fast evolution into longer wavelengths that can be approximated by $\lambda_d \approx \lambda_{max}[(f/3)^2+1]/2$. Also, even though the field will continue to be amplified until $v_A \sim v_{d,cr}$, the nonlinear growth will be slower than in the linear regime.

In the third part, we include the back-reaction on the CRs and find that the CR deflection by the amplified field constitutes another possible saturation mechanism. We tested this effect for a semi-isotropic distribution of monoenergetic CRs propagating at $v_{d,cr} = 0.5c$ with respect to the upstream medium, which would be appropriate for the most energetic CRs that escape from SNRs. We find that the field is amplified until the Larmor radii of the CRs becomes approximately equal to the size of the dominant magnetic fluctuations. When that happens, the CRs get strongly deflected by the magnetic field, which decreases their current and stops the growth of the field. Ignoring the migration to longer wavelengths, for a generic CR momentum distribution with $v_{d,cr} < c$, saturation due to CR deflection will happen when $f \approx (\Gamma_{cr}/4\pi)(c^2/v_{A,0}^2)(n_{cr}/n_i)(v_{d,cr}/c)$, where $\Gamma_{cr}$ is the typical Lorentz factor of current-carrying CR. On the other hand, saturation due to plasma acceleration to $v_{d,cr}$ velocity occurs when $f \approx v_{d,cr}/v_{A,0}$. Thus, the deflection of CRs will dominate if the CR energy density is such that $\Gamma_{cr}n_{cr} < 4\pi n_iv_{A,0}/c$. If the migration to longer wavelengths is included, saturation due to CR deflection would happen at even smaller magnetic amplification.

In the upstream medium of SNR forward shocks, we expect the CR energy density to be low enough so that the maximum CRCD amplification is determined by the back-reaction on the CRs. In order to make an estimate of typical magnetic amplification in these environments, let us consider a piece of upstream whose distance from the shock is such that it can only feel the most energetic CRs that escape from the remnant. We use only the most energetic particles because they are the only ones whose Larmor radii are much larger than the typical wavelength of the CRCD waves, which is an essential condition of the instability. Also, our estimate is based on the following assumptions. First, all the escaping particles have positive charge, which is reasonable considering the much shorter cooling time of electrons compared to ions, and that ions are presumably more efficiently injected into the acceleration process in shocks. Second, we assume that the escaping CRs have the same energy, $E_{esc}$, which is roughly the minimum energy required for them to run away from the remnant. Third, there is a fixed ratio, $\eta_{esc} \equiv F_{E,cr}/(\rho v_{sh}^3/2)$, between the flux of CR energy emitted by the shock, $F_{E,cr}$, and the flux of energy coming from the upstream medium as seen from the frame of the shock, $\rho v_{sh}^3/2$, where $v_{sh}$ is the shock velocity and $\rho$ is the mass density of the upstream plasma. Finally, we assume a plane geometry and that all the ions are protons. Under these conditions, the time scale of growth of the instability, $\gamma_{max}^{-1}$, is  
\begin{eqnarray}
\begin{array}{rrrl}
\gamma_{max}^{-1} & \approx & 50 \bigg(\frac{E_{esc}}{10^{15}\textrm{eV}}\bigg) \bigg(\frac{10^4\textrm{km/sec}}{v_{sh}}\bigg)^3 & \\ && \bigg(\frac{0.05}{\eta_{esc}}\bigg)\bigg(\frac{\textrm{cm}^{-3}}{n_i}\bigg)^{\frac{1}{2}}& \textrm{years,} 
\end{array}
\label{eq:timescale}
\end{eqnarray}
and the initial length scale of maximum growth is
\begin{eqnarray}
\begin{array}{rrrl}
\lambda_{max} & \approx & 3 \times 10^{-3} \bigg(\frac{v_{A,0}}{10\textrm{km/sec}}\bigg)\bigg(\frac{E_{esc}}{10^{15}\textrm{eV}}\bigg) & \\ && \bigg(\frac{10^4\textrm{km/sec}}{v_{sh}}\bigg)^3\bigg(\frac{0.05}{\eta_{esc}}\bigg)\bigg(\frac{\textrm{cm}^{-3}}{n_i}\bigg)^{\frac{1}{2}} & \textrm{pc.} 
\end{array}
\label{eq:lengthscale}
\end{eqnarray}
The ratio $\gamma_{max}/\omega_{c,i}$ is
\begin{eqnarray}
\begin{array}{rrr}
\gamma_{max}/\omega_{c,i} & \approx & 1.4 \times 10^{-8} \bigg(\frac{10\textrm{km/sec}}{v_{A,0}}\bigg) \bigg(\frac{10^{15}\textrm{eV}}{E_{esc}}\bigg) \\ && \bigg(\frac{v_{sh}}{10^4\textrm{km/sec}}\bigg)^3\bigg(\frac{\eta_{esc}}{0.05}\bigg), 
\end{array}
\label{eq:ratio}
\end{eqnarray}
which confirms that in the case of SNRs the CRCD instability will not be affected by the Weibel-like filamentation studied in \S\ref{sec:magnetization}. Considering the migration into longer wavelengths given by Equation (\ref{eq:wavelength}), the amplification factor, $f$, in SNRs will satisfy
\begin{equation}
f[(f/3)^2 + 1] \approx 130\bigg(\frac{10 \textrm{km/sec}}{v_{A,0}}\bigg)^2 \bigg(\frac{\eta_{esc}}{0.05}\bigg)\bigg(\frac{v_{sh}}{10^4\textrm{km/sec}}\bigg)^3,
\label{eq:estimate}
\end{equation}
which, for typical parameters would imply $f \approx 10$. We see that, even though run-away CRs can significantly amplify the ambient magnetic field, the upstream amplification alone is not enough to explain the factors of $\sim100$ inferred from observations of forward shocks in young SNRs \citep{VolkEtAl05, Ballet06, UchiyamaEtAl07}. Also, note that, if $E_{esc} = 10^{15}\textrm{eV}$ and $n_i = 1\textrm{cm}^{-3}$, the distance swept by the shock in a time $\gamma_{max}^{-1}$ is $v_{sh}\gamma_{max}^{-1} \approx 0.5 \textrm{pc}$, which is comparable to the typical size of a SNR. This means that the advection of the upstream fluid into the shock may happen faster than the growth of the field, and may put further restrictions on the amplification. 

It has been suggested that a further CRCD amplification could be provided by the current of lower energy CRs that are confined closer to the shock and move diffusively at drift velocity $v_{d,cr} \sim v_{sh}$ \citep{Bell04}. 
We believe, however, that this possibility requires a more detailed study. Such lower energy CRs can be magnetized in the sense that their Larmor radii are smaller than the typical size of magnetic fluctuations, $\lambda_d$, violating the conditions for the CRCD instability.
Although on large scales these CRs will still produce a current of magnitude $\sim e n_{cr}v_{sh}$ parallel to the shock normal, on scales of the CRCD wavelength the local CR current may get significantly affected by the amplified field because of the deflection of CRs. Field amplification may then proceed in essentially different way. 

The magnetization of CRs could be less of an issue if the wavelength of the instability due to low energy CRs is shorter than the CR Larmor radius. Indeed, since lower energy CRs are more numerous than the most energetic ones, their larger current will generate shorter CRCD waves (remember that $\lambda_{max} = B_0c/J_{cr}$). However, the CRCD turbulence generated further upstream by the highest energy CRs may modify the condition $\vec{J}_{cr} \parallel \vec{B}_0$ and may suppress the growth of the small wavelength modes closer to the shock. This suggests that other non-linear mechanisms, such as the cyclotron resonance of CRs with Alfv\'en waves \citep{KulsrudEtAl69, McKenzieEtAl82}, may still be important components in the amplification of the field. The full effect of the low-energy CR contribution needs to be investigated using a fully kinetic treatment of CRs that includes their reacceleration by the shock, the presence of pre-existing turbulence, and the eventual contribution of CR electrons to the cancelation of ion current. 

Although we have applied our results only to the non-relativistic case of SNRs ($v_{d,cr} < c$), the CRCD instability may also play an important role in relativistic shocks in jets and Gamma Ray Bursts, where $v_{d,cr} \cong c$. Our simulations show that, at constant CR current, the evolution of the instability is the same as in the non-relativistic case. In particular, the intrinsic saturation criterion due to plasma acceleration is valid, implying a maximum magnetic fiel such that $v_A \sim c$. However, if the back-reaction on the CRs is considered, the non-linear evolution of the field and the saturation due to CR deflection may be dominated by CR beam filamentation \citep{RiquelmeSp08}. Also, since in the upstream of GRB shocks the density of CRs might be close to the density of upstream ions \citep{CouchEtAl08}, the magnetization requirement ($\gamma_{max} \ll \omega_{c,i}$) may not be satisfied in these environments. In this case, a non-linear magnetic amplification is still expected, but through an instability that is characterized by Weibel-like filamentation of the plasma and whose properties may be different to the CRCD instability described here \citep{NiemiecEtAl08}. Detailed analysis of the relativistic shock case in application to GRBs will be presented elsewhere. 

In conclusion, we have shown that the CRCD instability is a viable mechanism for the non-linear amplification of magnetic field upstream of both non-relativistic and relativistic shocks, and that it can provide an efficient scattering mechanism for CRs in these environments.

\acknowledgments

This research is supported by NSF grant AST-0807381 and US-Israel Binational Science Foundation grant 2006095. A.S. acknowledges the support from Alfred P. Sloan Foundation fellowship. We thank Yury Lyubarsky and Ehud Nakar for useful discussions.

\appendix

\section{A) CRCD waves dispersion relation}\label{app:analytic}

In this appendix we calculate a dispersion relation for the CRCD waves for the case where $\vec{B}_{0}$, $\vec{J}_{cr}$, and the wave vector of the electromagnetic mode, $\vec{k}$,  are all parallel and point along the $\hat{x}$ axis. We will separate the fields and currents into components that are transverse and parallel to $\hat{x}$, and will identify them with the subscripts $``tr"$ (standing for ``transverse") and $``x"$, respectively. Thus, the magnetic field perpendicular to $\hat{x}$ will be given by
\begin{eqnarray}
\vec{B}_{tr} = \textrm{Re}\{ B_{0} (i\hat{y} + \hat{z}) e^{i(kx-\phi(t))+\gamma t}  \},
\label{eq:b}
\end{eqnarray}
which corresponds to a right-handed polarized wave, where the phase $\phi(t)$ is an unknown function of time, $t$, the growth rate $\gamma>0$ is a constant, and $B_{0}$ is the magnitude of the initial background field $\vec{B}_{0}$. Note that the time is chosen so that the wave is in the linear regime for $t<0$.

Then, from the Ampere's and Faraday's laws, we get that the electric field and the current perpendicular to $\hat{x}$ are given by
\begin{eqnarray}
\vec{E}_{tr} = \textrm{Re}\bigg\{ \frac{B_{0}(\omega + i\gamma)}{kc} (\hat{y}-i\hat{z}) e^{i(kx-\phi(t))+\gamma t}  \bigg\},
\label{eq:e}
\end{eqnarray}
and
\begin{eqnarray}
\vec{J}_{tr} = \textrm{Re}\bigg\{ \frac{B_{0}}{4 \pi kc} (k^2c^2 - (\omega + i\gamma)^2 - i\dot{\omega}) (-i\hat{y}-\hat{z}) e^{i(kx-\phi(t))+\gamma t}  \bigg\},
\label{eq:jotaperp}
\end{eqnarray}
where $\omega$ and $\dot{\omega}$ are the time and second time derivative of $\phi$, respectively.
In order to obtain a dispersion relation, we need another expression connecting $\vec{J}_{tr}$ with $\vec{E}_{tr}$ and $\vec{B}_{tr}$. We find it by making the following assumptions. First, the thermal velocities of the particles in the background plasma will not give rise to any significant drift velocity. Second, we will assume that $\gamma$, $|d(kx + \phi(t))/dt|$ and $(d^n[\omega]/dt^n)/(d^{(n-1)}[\omega]/dt^{(n-1)})$ are constant and much smaller than $|q_jB/m_jc| = \omega_{c,j}$, where $\omega_{c,j}$, $q_j$, and $m_j$ are the cyclotron frequency, the charge, and the mass of the $j$ species, respectively. (We will see at the end of this appendix that, in order to satisfy the last three conditions, we need $v_{A,0}/c \gg (n_{CR}/n_i)(v_{d,cr}/c)$, where $v_{A,0}$ is the initial Alfv\'en velocity of the plasma.) Third,  the electric and magnetic fields are perpendicular, which is a reasonable assumption in the case of a quasineutral plasma. And finally, $v_{d,cr} \gg v_{A}$, where $v_{A}$ is the Alfv\'en velocity of the plasma and $v_{d,cr}$ is the drift velocity of the CRs. Considering this, given $\vec{E}_{tr}$ and $\vec{B}_{tr}$, we can find $\vec{J}_{tr}$ as follows. 

If a particle $``j"$ experiences electric and magnetic fields, $\vec{E}$ and $\vec{B}$, its velocity perpendicular to $\vec{B}$ has two components, $\vec{v}_{g,j}$ and $\vec{v}_{d,j}$, that satisfy the equations $d\vec{v}_{g,j}/dt = (q_j/m_j)(\vec{v}_{g,j}/c) \times \vec{B}$ and $d\vec{v}_{d,j}/dt = (q_j/m_j)(\vec{E} + \vec{v}_{d,j}/c \times \vec{B})$. In the case of constant and uniform $\vec{E}$ and $\vec{B}$, $\vec{v}_{g,j}$ represents the classical gyration around $\vec{B}$, while $\vec{v}_{d,j}$ corresponds to the drift of the particle, which is $\vec{v}_{d,j} = c(\vec{E}\times\vec{B})/B^2$. 

When the fields change both in time and space, we can still decompose the velocity perpendicular to the field into $\vec{v}_{g,j} + \vec{v}_{d,j}$ [again, satisfying $d\vec{v}_{g,j}/dt = (q_j/m_j)(\vec{v}_{g,j}/c) \times \vec{B}$ and $d\vec{v}_{d,j}/dt = (q_j/m_j)(\vec{E} + \vec{v}_{d,j}/c \times \vec{B})$]. In this case, the space and time variations of $\vec{B}$ can also produce drift velocities due to the $\vec{v}_{g,j}$ motion. The space variations will give rise to a drift due to the curvature of the magnetic field lines (curvature drift). The curvature drift velocity, however, is of the order of the thermal speed of the particles times the ratio between their Larmor radii and the curvature radius of the lines. The time variations of the field, on the other hand, can also give rise to drift velocities. To first order, these velocities will also be proportional to the thermal speed of the particles times the ratio between the rate of change of the field (determined by the quantities $\gamma$ and $|d(kx+\phi(t))/dt|$) and $\omega_{c,j}$. We will neglect these possible drift velocities using our first assumption that the thermal velocities of the particles are low enough not to produce any important drift velocity.

On the other hand, in the case of a non-uniform and time-changing fields, the $\vec{v}_{d,j}$ velocity is given by the series, 
\begin{eqnarray}
\vec{v}_{d,j} = \sum_{n=0}^{\infty}\vec{v}_{d,j}^{(n)},
\end{eqnarray}
where 
\begin{eqnarray}
\begin{array}{lll}
\vec{v}_{d,j}^{(0)} = \frac{(\vec{E}\times\vec{B})c}{B^2} & \textrm{and} & \frac{q_j\vec{v}_{d,j}^{(n)}}{m_jc}\times \vec{B} = \frac{d\vec{v}_{d,j}^{(n-1)}}{dt},
\end{array}
\label{eq:velocities}
\end{eqnarray}
for $n=1,2,...,\infty$. 
We see from Equations (\ref{eq:b}), (\ref{eq:e}), and (\ref{eq:velocities}) that $|\vec{v}_{d,j}^{(n)}| \ll |\vec{v}_{d,j}^{(n-1)}|$ as long as $\gamma$, $|d(kx + \phi(t))/dt|$, and $(d^n[\omega]/dt^n)/(d^{(n-1)}[\omega]/dt^{(n-1)})$ are constant and much smaller than $|q_jB/m_jc| = \omega_{c,j}$, which is our second assumption. Notice that, even if $|\vec{v}_{d,j}^{(1)}| \ll |\vec{v}_{d,j}^{(0)}|$, the currents produced by these two velocities can be comparably important. This is because $\vec{v}_{d,j}^{(0)}$ is independent of $m_j$ and $q_j$, thus it has the same value for ions and electrons (so from now we will just drop the subscript $``j"$ and will refer to this velocity as simply $\vec{v}_d^{(0)}$). Thus, when considering both species, the current produced by $\vec{v}_{d}^{(0)}$, $\vec{J}_d^{(0)}$, will be due to the tiny excess of electrons in the background required to compensate the CRs charge, so it will be proportional to $n_{cr}$. On the other hand, since $\vec{v}_{d,j}^{(1)}$ is proportional to $m_j/q_j$, it will be much larger for ions than for electrons. So the corresponding $\vec{J}_d^{(1)}$ will be proportional to the total density of ions in the background, $n_i$, which is typically much larger than $n_{cr}$. Since $\vec{v}_{d,j}^{(n)}$ for $ n \ge 2$ will also affect mainly the ions, their contribution to the current in the plasma will be much smaller than the one of $\vec{v}_{d,j}^{(1)}$
provided that $\vec{v}_{d,j}^{(n)} \ll \vec{v}_{d,j}^{(n-1)}$, so we will just neglect them. 
Thus the currents $\vec{J}_d^{(0)}$ and $\vec{J}_d^{(1)}$ can be calculated considering Equations (\ref{eq:b}), (\ref{eq:e}), and (\ref{eq:velocities}), finding that
\begin{equation} 
\vec{J}_d^{(0)} = \frac{J_{cr}}{kv_{d,cr}}\bigg[
-f^2\frac{\omega}{1+f^2}\hat{i} + \textrm{Re}\bigg\{ \bigg(-i\gamma -
\frac{\omega}{1+f^2}\bigg)e^{i(kx - \phi(t)) + \gamma
t}(-i\hat{y}-\hat{z}) \bigg\}\bigg],
\label{eq:jotazero}
\end{equation}
and
\begin{eqnarray}
\begin{array}{llr}
\vec{J}_d^{(1)} & = & \frac{B_{0}}{4\pi
kc}\frac{c^2}{v_A^2} \bigg[f^2\bigg(-\gamma^2 + \bigg(\frac{\omega}{1+f^2}\bigg)^2\bigg)
\hat{i} + \textrm{Re}\bigg\{\bigg(2\gamma \frac{d\phi}{dt}i
\Big(\frac{1}{1+f^2}+\frac{\dot{\omega}}{2\gamma \omega}\Big) \\
& & -\gamma^2 + \Big(\frac{\omega}{1+f^2}\Big)^2\bigg) e^{i(kx - \phi(t)) + \gamma t}(-i\hat{y}-\hat{z})
\bigg\}\bigg], 
\end{array}
\label{eq:jotauno}
\end{eqnarray} 
where we have defined the field amplification factor $f \equiv B_{tr}/B_{0}$. \newline
This way we have calculated all the currents in the plasma that are perpendicular to $\vec{B}$, but we still have to determine the ones that are parallel to $\vec{B}$. We do that using our third assumption, $\vec{E} \perp \vec{B}$, which implies that
\begin{eqnarray}
E_x=-(\vec{E}_{tr}\cdot\vec{B}_{tr})/B_x.
\label{eq:perpendicularity}
\end{eqnarray}
Using the Ampere's law and the fact that in a one dimensional problem $(\nabla \times
\vec{B})_x = 0$, we have that 
\begin{equation}
\frac{\partial E_x}{\partial t} = -4\pi(J_{cr} + J_{d,x}^{(0)} +
J_{d,x}^{(1)} + J_{\parallel,x}),
\label{eq:jotapax}
\end{equation}
where $J_{\parallel,x}$ is the $x$ component of the plasma current parallel to $\vec{B}$, $\vec{J}_{\parallel}$. Given this, the component of $\vec{J}_{\parallel}$ perpendicular to $\hat{x}$ is just
\begin{equation}
\vec{J}_{\parallel,tr} = J_{\parallel,x} \frac{\vec{B}_{tr}}{B_{0}}.
\label{eq:jotapape}
\end{equation}
Then, using Equations (\ref{eq:b}), (\ref{eq:e}), (\ref{eq:perpendicularity}), (\ref{eq:jotapax}) and (\ref{eq:jotapape}) we get that
\begin{equation}
\vec{J}_{\parallel,tr} =  \textrm{Re}\bigg\{
\bigg(J_{cr}\bigg(1-\frac{f^2}{kv_{d,cr}}\frac{\omega }{1+f^2}\bigg) + \frac{B_{0}f^2}{4\pi
kc}\bigg(-\gamma^2\bigg(2 + \frac{c^2}{v_A^2}\bigg) + \frac{c^2}{
v_A^2}\bigg(\frac{\omega}{1+f^2}\bigg)^2\bigg)\bigg)e^{i(kx - \phi(t)) + \gamma
t}(-i\hat{y}-\hat{z})\bigg\}.
\label{eq:jotapa}
\end{equation} 
Now we have the expressions for all the components of the current perpendicular
to $\hat{x}$, $\vec{J}_{tr}$, so we can use Equation (\ref{eq:jotaperp}) to find the dispersion
relation,
\begin{eqnarray}
\begin{array}{llr}
k^2c^2 - (\omega + i\gamma)^2 - i \dot{\omega} & = & \frac{4\pi
J_{cr}}{B_{0}}\Big(kc-\frac{c}{v_{d,cr}}(\omega + i\gamma)\Big) -\gamma^2 \Big(2f^2 + (1+f^2)\frac{c^2}{v_A^2}\Big)
\\ %\nonumber
&& + \frac{c^2}{v_A^2}\frac{\omega^2}{1+f^2} + i2\gamma \frac{c^2
}{v_A^2}\omega\Big(\frac{1}{1+f^2} +
\frac{\dot{\omega}}{2\gamma\omega}\Big).
\end{array}
\label{eq:dispersion}
\end{eqnarray}

From this derivation we can also obtain an estimate of the plasma velocities due to the CRCD waves. We know that the motion of particles perpendicular to $\vec{B}$ is dominated by $\vec{v}_d^{(0)} = -\vec{J}_d^{(0)}/n_{cr}$ (since $\vec{v}_d^{(0)}$ affects ions and electrons in the same way and $n_e - n_i = n_{cr}$), and that the motion of particles parallel to $\vec{B}$ is mainly given by electrons moving at $\vec{v}_{\parallel} = -\vec{J}_{\parallel}/n_e$. Thus, by looking at the expressions for $\vec{J}_d^{(0)}$ and $\vec{J}_{\parallel}$ given by Equations (\ref{eq:jotazero}), (\ref{eq:jotapape}), and (\ref{eq:jotapa}), we find that the dominant plasma motion will be given by $\vec{v}_d^{(0)}$ and will imply a velocity of ions and electrons that can be decomposed into a component along $x$, $v_{x,an} \approx f^2v_{A,0}^2/v_{d,cr}$ (where the subscript $``an"$ stands for $``analitic"$), and a transverse component, $v_{tr,an} \approx fv_{A,0}$, that is always perpendicular to $\vec{B}_{tr}$ (and to $\hat{x}$).

In \S \ref{sec:analytical} we use the dispersion relation given by Equation (\ref{eq:dispersion}) to calculate the wavenumber, $k_{max}$, and growth rate, $\gamma_{max}$, of the fastest growing mode, as well as $\omega$ as a function of the amplitude of the wave. We will use these results here in order to check the consistency of assuming that $\gamma$, $|d(kx + \phi(t))/dt|$, and $(d^n[\omega]/dt^n)/(d^{(n-1)}[\omega]/dt^{(n-1)})$ are constant and much smaller than $\omega_{c,i}$, which are necessary for neglecting $|\vec{v}_d^{(n)}|$ when $n \ge 2$. We know from Equation (\ref{eq:phasevelocity}) that when $v_A \ll v_{d,cr}$, $\omega \approx k_{max}v_A^2/v_{d,cr}$. It means that $(d^{(n)}[\omega]/dt^n)/ (d^{(n-1)}[\omega]/dt^{(n-1)})  \approx 2\gamma$ (except when $n=1$ and $f \ll 1$, in which case $(d^{(n)}[\omega]/dt^n)/ (d^{(n-1)}[\omega]/dt^{(n-1)})  \ll \gamma$). So, in order to neglect $|\vec{v}_d^{(n)}|$ for $n \ge 2$, we only require $\gamma$ and $|d(kx + \phi(t))/dt|$ to be constant and much smaller than $\omega_{c,i}$. It is possible to show from Equation (\ref{eq:growthrate}) that the first condition, which is equivalent to $\gamma_{max} \ll \omega_{c,i}$, where $\gamma_{max}$ is given by Equation (\ref{eq:growthrate}), is satified if $v_{A,0}/c \gg (n_{CR}/n_i)(v_{d,cr}/c)$. From Equation (\ref{eq:phasevelocity}) we see that, if we approximate $dx/dt \approx v_{d,x}^{(0)}$, $d(kx + \phi(t))/dt$ becomes approximately equal to $k_{max}c (v_{A,0}/c)^2 \approx \gamma_{max}(v_{A,0}/c)$, which is also constant and much smaller than $\gamma_{max}$. This way we see that our analytical results are valid if the plasma is well magnetized in the sense that $\gamma_{max} \ll \omega_{c,i}$, which is equivalent to $v_{A,0} \gg (n_{CR}/n_i) v_{d,cr}$. Using two-dimensional PIC simulations, we show in \S\ref{sec:magnetization} that this condition is actually a requirement for the CRCD not to be quenched by the Weible-like filamentation.

\section{B) Multidimensional evolution: two-dimensional simulations}\label{sec:2d}

\begin{figure*}
\centering\includegraphics[width=13cm]{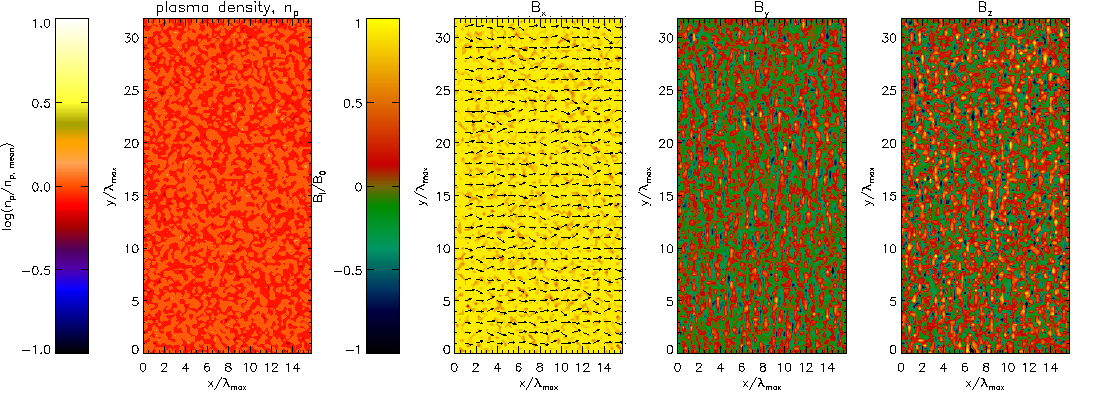}
\caption{Plasma density, $n_p$, and the three components of the magnetic field: $B_x$, $B_y$, and $B_z$ at $t\gamma_{max} = 6$ for two-dimensional run I4 of Table \ref{table:interference}. The arrows on the $B_x$ panel show the direction of the magnetic field projected on the $x-y$ plane. This stage shows the CRCD waves growing at wavelengths $\sim \lambda_{max}$ in their linear stage ($B_{tr} \ll B_0$).} \label{fg:shocks1}
\end{figure*}
\begin{figure*}
\centering\includegraphics[width=13cm]{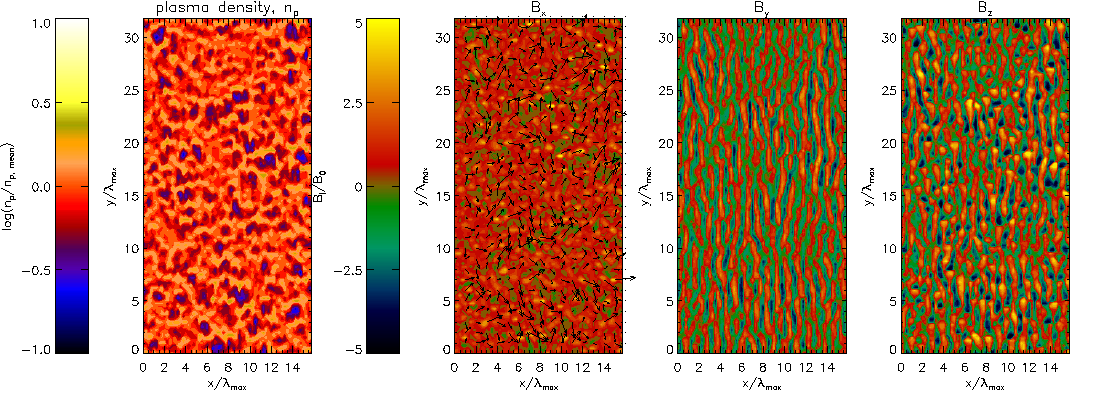}
\caption{Same as in Fig. \ref{fg:shocks1}, but at $t\gamma_{max} = 9$ (when $B_{tr} \sim B_0$). Prominent density fluctuations ($\Delta n_p/n_p \sim 10$) on scales of $\sim \lambda_{max}$ are starting to develop. Also, the CRCD waves are starting to get distorted due to the growing importance of the turbulent motions in the plasma.} \label{fg:shocks2}
\end{figure*}
\begin{figure*}
\centering\includegraphics[width=13cm]{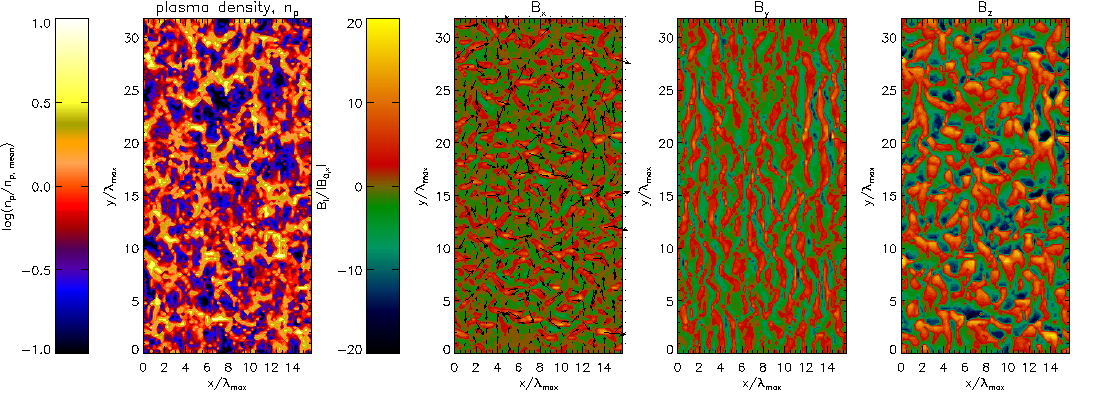}
\caption{Same as in Fig. \ref{fg:shocks1}, but at $t\gamma_{max} = 11$ (when $B_{tr} \gg B_0$). The amplitude of the density fluctuations is still prominent ($\Delta n_p/n_p \sim 10$), but their length scale is larger, compared to the $t\gamma_{max} = 9$ case (Fig. \ref{fg:shocks2}). The magnetic fluctuations have also grown in size, and a clear differentiation between $B_y$ and $B_z$, peculiar to two-dimensional runs, can be observed.} \label{fg:shocks3}
\end{figure*}
We saw in \S \ref{sec:threed} that, when multidimensional effects are considered, the dominant wavelength of the CRCD instability, $\lambda_d$, grows according to Equation (\ref{eq:wavelength}). This makes it numerically expensive to run three-dimensional simulations that could amplify the field substantially without making $\lambda_d$ too close to the size of the simulation box, $L$. In order to overcome this difficulty, in this section we present the results of two-dimensional simulations whose $L$ is always bigger than $\lambda_d$. Despite some artifacts related to the two-dimensional geometry, these simulations help us confirm the main results obtained from the three-dimensional analysis presented above. 

Figures \ref{fg:shocks1}, \ref{fg:shocks2}, and \ref{fg:shocks3} show the results at three different times ($t\gamma_{max} = 6$, 9, and 11, respectively) for one of the simulations (run I4 of Table \ref{table:interference}), which corresponds to CRs drifting at the speed of light and without considering their back-reaction. We see that initially the instability is produced independently in different regions of the simulation box (as seen in Fig. \ref{fg:shocks1}). In this linear stage of evolution, the waves produced in adjacent regions of space seem to grow without interfering with each other. However, when the waves become non-linear, strong density fluctuations appear on scales of a few $\lambda_{max}$ (as shown in Fig. \ref{fg:shocks2}). The beginning of this stage is shown in Fig. \ref{fg:shocks2}. It is also apparent from Fig. \ref{fg:shocks3} that the magnetic fluctuations get distorted and evolve into larger scales right after the density fluctuations and turbulence form.

\begin{figure}
\centering\includegraphics[width=7.75cm]{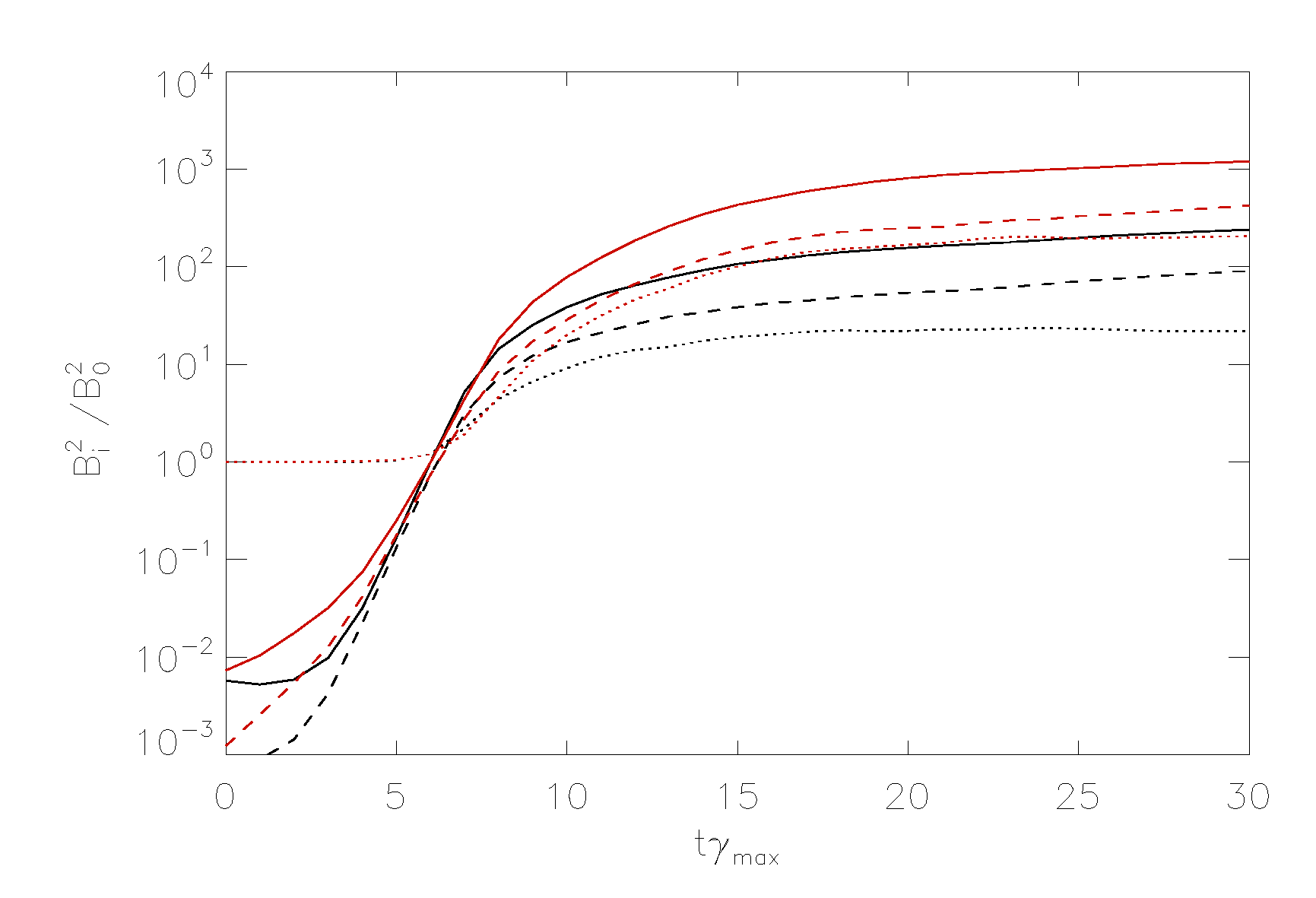}
\caption{The three components of the magnetic energy as a function of time for the two-dimensional runs I5 and I6 of Table \ref{table:interference}, represented by black and red lines, respectively. The $x$, $y$, and $z$ components are shown using dotted, dashed, and solid lines, respectively. Time is normalized in terms of the $\gamma_{max}$ of each simulation. The differentiation between the $y$ and $z$ components of the field as well as the departure from exponential growth after $B_{tr} \sim B_0$ can be seen for both runs. Saturation still happens when $v_A \sim v_{d,cr}$.} \label{fg:departure}
\end{figure}

The formation of density fluctuations also affects the growth rate of the instability. Fig. \ref{fg:departure} presents the magnetic energy evolution for the two-dimensional simulations I5 and I6, whose $v_{A,0}/c = 1/10$ and $1/50$, respectively. The rest of their numerical parameters are specified in Table \ref{table:interference}. We see that, as in the three-dimensional case, the exponential growth stops shortly after $B_{tr} \sim B_0$. After that, the CRCD instability grows at a lower rate, reaching saturation when $v_A \sim v_{d,cr}$. This result had already been obtained in the three-dimensional case, but in this case we allow the instability to evolve into larger scales as the magnetic field grows.

In two dimensions, the formation of density fluctuations produces a clear differentiation between the $y$ and $z$ components of the field (as can be seen in Figs. \ref{fg:shocks3} and \ref{fg:departure}). This is because in the low density regions the plasma cannot generate the return current necessary to compensate $\vec{J}_{cr}$. Thus, the uncompensated CR current produces a ``toroidal" magnetic field around the underdense regions that, in the two-dimensional case, manifests itself as an amplification of the out of the plane component of the field, $B_z$. Even though, as seen in Fig. \ref{fg:shocks3}, both the ``toroidal" and the CRCD field coexist, the two-dimensional simulations can still give us information about the point when the CRCD instability stops amplifying the field.

\end{document}